\documentclass[twocolumn]{aastex631}

\def\OI{[\mbox{O\,{\sc i}}]$\lambda 6300$}
\def\OIII{[\mbox{O\,{\sc iii}}]$\lambda 5007$}

\def\OIIIab{[\mbox{O\,{\sc iii}}]$\lambda\lambda 4959,5007$}

\def\SII{[\mbox{S\,{\sc ii}}]$\lambda \lambda 6717,6731$}
\def\NII{[\mbox{N\,{\sc ii}}]$\lambda 6584$}
\def\NIIb{[\mbox{N\,{\sc ii}}]$\lambda 6584$}

\def\NIIab{[\mbox{N\,{\sc  ii}}]$\lambda \lambda 6547,6584$}

\def\OIIa{[\mbox{O{\sc ii}}]$\lambda 3727$} 

\def\HeII{{He{\sc ii}}$\lambda 4686$}

\def\OI{[\mbox{O{\sc i}}]$\lambda 6300$}

\def\Ha{{H$\alpha$}}
\def\Hb{{H$\beta$}}
\def\Hg{{H$\gamma$}}
\def\Hd{{H$\delta$}}

\def\NIIHa{[\mbox{N\,{\sc ii}}]$\lambda 6583$/H$\alpha$}

\def\OIIIHb{[\mbox{O\,{\sc iii}}]$\lambda 5007$/H$\beta$}

\def\LOIII{$L[\mbox{O\,{\sc iii}}]$}

\def\LOIIIs4{$L[\mbox{O\,{\sc iii}}]$/$\sigma^4$}

\def\kms{${\rm km}~{\rm s}^{-1}$}
\newcommand{\ergcms}	{\ifmmode {\rm erg\,cm}^{-2}\,{\rm s}^{-1} \else erg\,cm$^{-2}$\,s$^{-1}$\fi}

\definecolor{myblue}{RGB}{0, 100, 220}
\definecolor{myred}{RGB}{225, 0, 100}

\usepackage{gensymb}
\usepackage{graphicx}
\usepackage[figuresright]{rotating} 
\usepackage{url} 
\usepackage{multirow}
\usepackage{CJK}
\usepackage[nameinlink,capitalise]{cleveref}

\shorttitle{BASS DR2: Line measurements and AGN demographics}
\shortauthors{Oh et al.}

\begin{document}

\title{BASS XXIV: The BASS DR2 Spectroscopic Line Measurements and AGN Demographics}

\correspondingauthor{Kyuseok Oh}
\email{oh@kasi.re.kr}

\author[0000-0002-5037-951X]{Kyuseok Oh}
\altaffiliation{JSPS Fellow}
\affil{Korea Astronomy and Space Science Institute, Daedeokdae-ro 776, Yuseong-gu, Daejeon 34055, Republic of Korea}
\affil{Department of Astronomy, Kyoto University, Kitashirakawa-Oiwake-cho, Sakyo-ku, Kyoto 606-8502, Japan}

\author[0000-0002-7998-9581]{Michael J. Koss}
\affiliation{Eureka Scientific, 2452 Delmer Street Suite 100, Oakland, CA 94602-3017, USA}
\affiliation{Space Science Institute, 4750 Walnut Street, Suite 205, Boulder, Colorado 80301, USA}

\author[0000-0001-7821-6715]{Yoshihiro Ueda}
\affiliation{Department of Astronomy, Kyoto University, Kitashirakawa-Oiwake-cho, Sakyo-ku, Kyoto 606-8502, Japan}

\author[0000-0003-2686-9241]{Daniel Stern}
\affiliation{Jet Propulsion Laboratory, California Institute of Technology, 4800 Oak Grove Drive, MS 169-224, Pasadena, CA 91109, USA}

\author[0000-0001-5231-2645]{Claudio Ricci}
\affiliation{N\'ucleo de Astronom\'ia de la Facultad de Ingenier\'ia, Universidad Diego Portales, Av. Ej\'ercito Libertador 441, Santiago 22, Chile}
\affiliation{Kavli Institute for Astronomy and Astrophysics, Peking University, Beijing 100871, People's Republic of China}
 
\author[0000-0002-3683-7297]{Benny Trakhtenbrot}
\affiliation{School of Physics and Astronomy, Tel Aviv University, Tel Aviv 69978, Israel}

\author[0000-0003-2284-8603]{Meredith C. Powell}
\affiliation{Kavli Institute of Particle Astrophysics and Cosmology, Stanford University, 452 Lomita Mall, Stanford, CA 94305, USA}

\author[0000-0002-8760-6157]{Jakob S. den Brok}
\affiliation{Institute for Particle Physics and Astrophysics, ETH Z{\"u}rich, Wolfgang-Pauli-Strasse 27, CH-8093 Z{\"u}rich, Switzerland}
\affiliation{Argelander Institute for Astronomy, Auf dem H{\"u}gel 71, 53231, Bonn, Germany}

\author[0000-0003-3336-5498]{Isabella Lamperti}
\affiliation{Centro de Astrobiolog\'{i}a (CAB, CSIC-INTA), Departamento de Astrof\'{i}sica, Ctra. de Ajalvir Km. 4, 28850 Torrej\'{o}n de Ardoz, Madrid, Spain}

\author[0000-0002-7962-5446]{Richard Mushotzky}
\affiliation{Department of Astronomy, University of Maryland, College Park, MD 20742, USA}

\author[0000-0001-5742-5980]{Federica Ricci}
\affiliation{Dipartimento di Fisica e Astronomia, Universit\`{a} di Bologna, via Gobetti 93/2, 40129 Bologna, Italy}
\affiliation{INAF - Osservatorio di Astrofisica e Scienza dello Spazio di Bologna, via Gobetti 93/3, I-40129 Bologna, Italy}

\author[0000-0001-5481-8607]{Rudolf E. B\"{a}r}
\affiliation{Institute for Particle Physics and Astrophysics, ETH Z{\"u}rich, Wolfgang-Pauli-Strasse 27, CH-8093 Z{\"u}rich, Switzerland}

\author[0000-0003-0006-8681]{Alejandra F. Rojas}
\affiliation{Centro de Astronom\'{i}a (CITEVA), Universidad de Antofagasta, Avenida Angamos 601, Antofagasta, Chile}

\author[0000-0002-4377-903X]{Kohei Ichikawa}
\affil{Frontier Research Institute for Interdisciplinary Sciences, Tohoku University, Sendai 980-8578, Japan}
\affil{Astronomical Institute, Graduate School of Science, Tohoku University, 6-3 Aramaki, Aoba-ku, Sendai 980-8578, Japan}
\affil{Max-Planck-Institut f{\"u}r extraterrestrische Physik (MPE), Giessenbachstrasse 1, D-85748 Garching bei M{\"u}unchen, Germany}

\author[0000-0002-1321-1320]{Rog\'{e}rio Riffel}
\affiliation{Departamento de Astronomia, Instituto de F\'\i sica, Universidade Federal do Rio Grande do Sul, CP 15051, 91501-970, Porto Alegre, RS, Brazil}

\author[0000-0001-7568-6412]{Ezequiel Treister}
\affiliation{Instituto de Astrof{\'i}sica, Facultad de F{\'i}sica, Pontificia Universidad Cat{\'o}lica de Chile, Casilla 306, Santiago 22, Chile}

\author{Fiona Harrison}
\affiliation{Cahill Center for Astronomy and Astrophysics, California Institute of Technology, Pasadena, CA 91125, USA}

\author[0000-0002-0745-9792]{C. Megan Urry}
\affiliation{Yale Center for Astronomy \& Astrophysics and Department of Physics, Yale University, P.O. Box 2018120, New Haven, CT 06520-8120, USA}

\author[0000-0002-8686-8737]{Franz E. Bauer}
\affiliation{Instituto de Astrof\'{\i}sica  and Centro de Astroingenier{\'{\i}}a, Facultad de F\'{i}sica, Pontificia Universidad Cat\'{o}lica de Chile, Casilla 306, Santiago 22, Chile}
\affiliation{Millennium Institute of Astrophysics (MAS), Nuncio Monse{\~{n}}or S{\'{o}}tero Sanz 100, Providencia, Santiago, Chile}
\affiliation{Space Science Institute, 4750 Walnut Street, Suite 205, Boulder, Colorado 80301, USA}

\author[0000-0001-5464-0888]{Kevin Schawinski}
\affiliation{Modulos AG, Technoparkstrasse 1, CH-8005 Zurich, Switzerland}



\newcommand{\Ntotalobtained}{743} 
\newcommand{\NPalomar}{271}
\newcommand{\NXshooter}{169}
\newcommand{\NSDSS}{118}
\newcommand{\NduPont}{41}
\newcommand{\NsixDF}{36}
\newcommand{\NSOAR}{32}
\newcommand{\NCTIO}{21}
\newcommand{\NSAAO}{17}
\newcommand{\NVT}{38}
\newcommand{\NKPNO}{9}
\newcommand{\NGemini}{7}
\newcommand{\NKeck}{5}
\newcommand{\NPerkins}{4}
\newcommand{\NCassini}{4}
\newcommand{\NTNG}{2}
\newcommand{\NSPM}{2}
\newcommand{\NHJS}{1}
\newcommand{\NUH}{1}
\newcommand{\NESOMPI}{1}
\newcommand{\NCopernico}{1}
\newcommand{\NTillinghast}{1}

\newcommand{\NAGN}{858}
\newcommand{\Nspec}{1425} 
\newcommand{\NnonbeamedAGN}{746}

\definecolor{mycolor}{rgb}{0.858, 0.188, 0.478}
\definecolor{mycolor2}{rgb}{0.07, 0.04, 0.76}

\begin{abstract}
We present the second catalog and data release of optical spectral line measurements and AGN demographics of the BAT AGN Spectroscopic Survey, which focuses on the Swift-BAT hard X-ray detected AGNs. We use spectra from dedicated campaigns and publicly available archives to investigate spectral properties of most of the AGNs listed in the 70-month Swift-BAT all-sky catalog; specifically, $\Ntotalobtained$ of the $746$ unbeamed and unlensed AGNs ($99.6\%$). We find a good correspondence between the optical emission line widths and the hydrogen column density distributions using the X-ray spectra, with a clear dichotomy of AGN types for $N_{\rm H} = 10^{22}~{\rm cm}^{-2}$. Based on optical emission-line diagnostics, we show that $48\%$--$75\%$ of BAT AGNs are classified as Seyfert, depending on the choice of emission lines used in the diagnostics. The fraction of objects with upper limits on line emission varies from $6\%$ to $20\%$. Roughly $4\%$ of the BAT AGNs have lines too weak to be placed on the most commonly used diagnostic diagram, \OIII/\Hb\ versus \NII/\Ha, despite the high signal-to-noise ratio (S/N) of their spectra. This value increases to $35\%$ in the \OIII/\OIIa\ diagram, owing to difficulties in line detection. Compared to optically-selected narrow-line AGNs in the Sloan Digital Sky Survey, the BAT narrow-line AGNs have a higher rate of reddening/extinction, with ${\rm H}\alpha/{\rm H}\beta>5$ ($\sim36\%$), indicating that hard X-ray selection more effectively detects obscured AGNs from the underlying AGN population. Finally, we present a subpopulation of AGNs that feature complex broad-lines ($34\%$, 250/743) or double-peaked narrow emission lines ($2\%$, 17/743). 
\end{abstract}



\section{Introduction}
\label{sec:intro}

Obscuration due to dusty material in active galactic nuclei (AGNs) is known to cause selection bias across almost all spectra regimes \citep[e.g.,][]{Hickox:2018:625}. The obscuring medium is likely located in the innermost area of the AGN near the central supermassive black hole (SMBH), approximately $\sim 1 - 100$ pc,  and can absorb a large fraction of the radiation emitted from the soft X-ray ($<10$ keV) to the optical bands \citep[][]{RamosAlmeida:2017:679}. AGN unification models \citep{antonucci93,urry95} explain this as being caused by the torodial structure of the main absorbing region. When the dusty torus blocks the line of sight to the nucleus, radiation emitted from the broad-line region, located inside dust torus, cannot reach us.

The narrow-line region (NLR), which is on larger kpc scales outside the torus and thus considered to be less sensitive to torus obscuration, has been used extensively to explore the central structure of AGNs and their spectroscopic properties. For example, optical narrow emission lines of a sizeable sample of AGNs from massive spectroscopic surveys, such as the Sloan Digital Sky Survey (SDSS, \citealt{york00}) and the Mapping Nearby Galaxies at APO (MaNGA, \citealt{Bundy15}), were used to diagnose the physical state of the ionized gas that differentiates AGNs from non-active galaxies and/or star-forming activity \citep[i.e., using the so-called BPT diagram;][]{Baldwin81, Veilleux87, Kewley01, Kewley06, Schawinski07, Wylezalek18}.

\begin{figure*}
\centering
	\includegraphics[width=0.56\linewidth, angle=270]{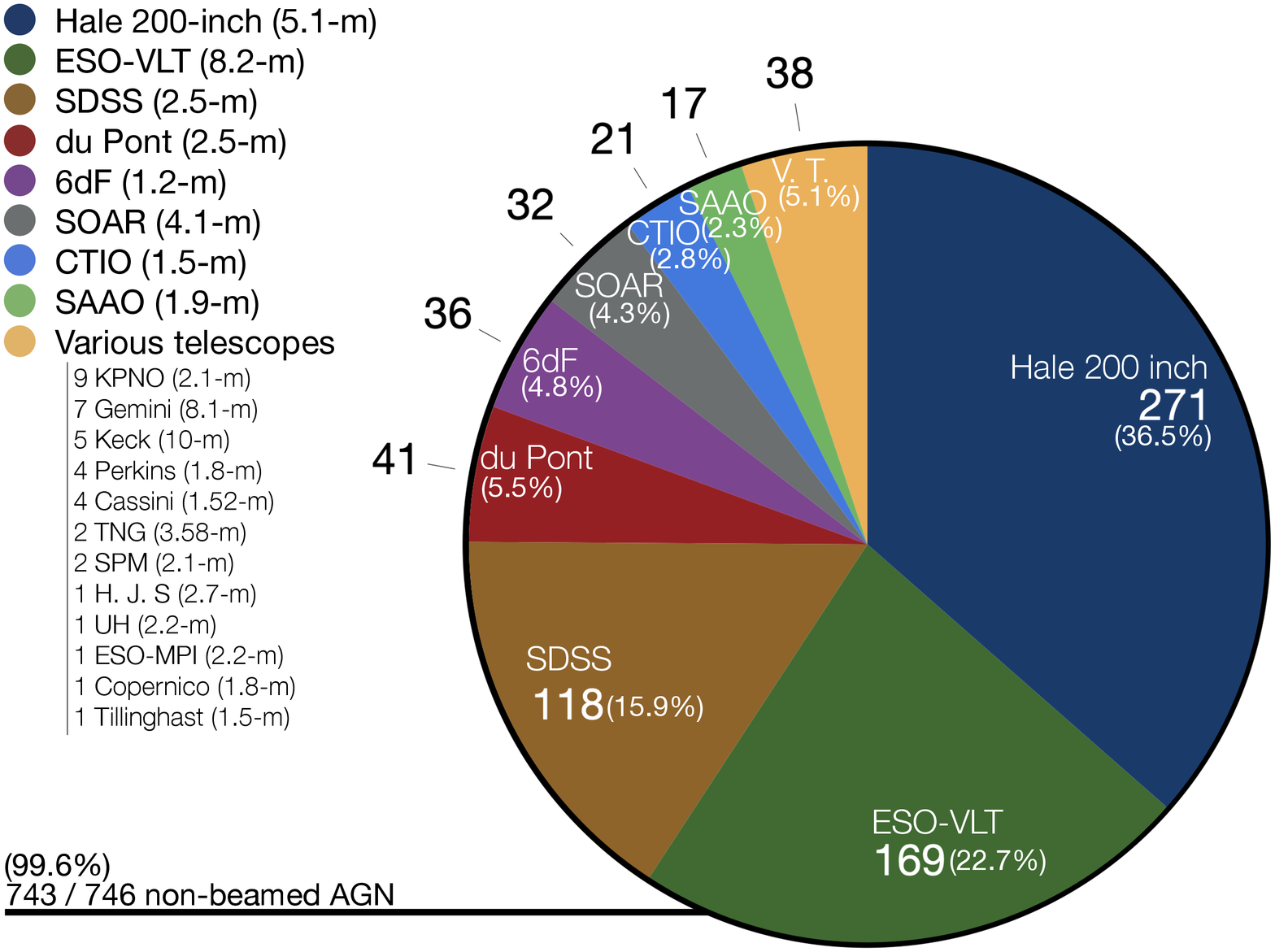} 
    \caption{Sources of \Ntotalobtained\ unique AGN spectra used for this study from the BASS DR2 and BASS DR1. Some AGN only BASS DR2 spectra in short wavelength high resolution setups not covering the full suite of emission lines, so DR1 data was used.  The BASS DR2 obtained spectra from targeted spectroscopic campaigns using the Hale 200-inch telescope ($N=\NPalomar$), ESO-VLT ($N=\NXshooter$), du Pont ($N=\NduPont$), SOAR ($N=\NSOAR$), and Keck ($N=\NKeck$). Spectra from publicly available surveys, such as SDSS ($N=\NSDSS$), 6dF ($N=\NsixDF$), and the BASS DR1 ($N=66$), are also used in this study.}
    \label{fig:fig1}
\end{figure*}

While BPT diagnostic diagrams allow large-scale surveys to identify narrow-line AGNs, optical spectroscopy often misses AGN signatures, either due to obscuration by dust in the host galaxy, or additional line emission (contamination) from star formation \citep{Elvis81, Iwasawa93, Comastri02, Goulding09}. Several studies have particularly noted that low-mass SMBHs are difficult to detect, owing to dilution from star formation \citep{Trump15, Cann19}. Furthermore, optical broad emission lines, which are often characteristic of AGNs, can be related to Type II supernovae \citep[SNe;][]{Filippenko97, Baldassare16}. Therefore, AGN selection using optical spectroscopy may miss significant populations of less powerful accreting SMBHs, particularly those hosted in star-forming galaxies.

By contrast, high-energy photons above 10 keV that are emitted in the vicinity of the AGN corona can penetrate the obscuring torus (e.g., X-ray photon count rates greater than $90\%$ for $\log N_{\rm H}<10^{23.5}\,{\rm cm}^{-2}$, \citealt{Ricci15, Koss16}); however, even X-rays can be biased at very high, Compton-thick columns (${\rm log}N_{\rm H}>10^{24}$~${\rm cm}^{-2}$), while other methods using the infrared or optical emission-line diagnostics may remain effective \citep{Georgantopoulos11, Goulding11, Severgnini12}.

 The Burst Alert Telescope \citep[BAT,][]{Barthelmy05} onboard the Swift satellite \citep[Neil Gehrels Swift Observatory;][]{Gehrels04} has been performing an ultra-hard X-ray all-sky survey at $14-195$ keV since 2005, and it has provided a set of the least-biased AGN source catalogs \citep{markwardt05, Tueller08, Tueller10, Baumgartner13, Oh18}. Compared with the earlier surveys at $13-180$ keV, conducted by HEAO 1 in the late 1970s \citep{Levine84}, the Swift-BAT survey significantly increased the number of known hard X-ray extragalactic sources by a factor of almost 25: the Swift-BAT 105-month survey has identified 1100 AGNs, of which 242 AGNs are newly identified \citep[][]{Oh18}. Most of these BAT AGNs are nearby ($ \langle z \rangle \simeq 0.05$) powerful AGNs that are as luminous as those detected by deep, narrow-field X-ray surveys that focus on high-redshift populations (\citealt{Brandt15}, and references therein). The BAT survey is also particularly useful as an accompaniment to the eROSITA mission \citep{Predehl21}, which is conducting an all-sky and much deeper survey in the softer X-ray regime ($0.5-10$ keV), where heavily obscured AGNs are harder to identify (e.g. \citealt{Koss16}).

However, despite the quantitative growth in the number of hard X-ray selected AGNs, comprehensive optical spectroscopic studies for a sizeable sample of these low-z AGNs (e.g. $N>50-100$) have been limited. The baseline of earlier optical spectroscopic follow-up studies was subsets of AGNs drawn from the 9-, 54-, and 70-month Swift-BAT survey catalogs \citep{Winter10, Parisi14, Ueda15, Rojas17, Marchesini19} and the 40-month catalog from the NuSTAR serendipitous survey \citep{Lansbury17}.

\begin{deluxetable*}{llrlll}[ht]
\tabletypesize{\small}
\tablecaption{Summary of instrumental setups for newly obtained spectra}
\label{tab:table1}
\tablewidth{0pt}
\tablehead{
\colhead{Telescope} & 
\colhead{Instrument} & 
\colhead{Total} &
\colhead{Grating} & 
\colhead{Slit Width[\arcsec]\tablenotemark{a}} &
\colhead{Resolution FWHM [\AA]\tablenotemark{b}}
}
\startdata
Hale 200 inch				&		DBSP				&		\NPalomar		&	600/316			&	1.5								&	4.1\\
ESO-VLT 					&		X-Shooter				&		\NXshooter		&	Echelle		&	1.6 (UVB), 1.5 (VIS)			&	1.3										 \\
du Pont 					&		B\&C					&		\NduPont			&	300			&	1								&	10.4										 \\
SOAR 					&		GOODMAN			&		\NSOAR			&	600, 400		&	1.2								&	3.8, 5.6										 \\
Keck						&		LRIS					&		\NKeck			&	600/400	&	1.0, 1.5		&      3.9, 4.6
\enddata
\tablenotetext{a}{For Palomar and X-Shooter, some smaller and larger slit widths were used for a few objects.}
\tablenotetext{b}{Resolution measured at 5000\,\AA.}
\tablecomments{A more detailed list of instrumental setups for the full BASS DR2 is provided in \citep{Koss_DR2_catalog}, here we provide a list for the spectra and telescopes used in this project. The instrumental setups of the optical spectra released in the DR1 are summarized in Table 1 of \citealt{Koss17}.}
\end{deluxetable*}

\begin{deluxetable*}{llllrlcccr}
\tabletypesize{\scriptsize}
\tablecaption{Optical spectra}
\tablewidth{0pt}
\tablecolumns{10}
\tablehead{
\colhead{\multirow{2}{*}{ID\tablenotemark{a}}} & 
\colhead{\multirow{2}{*}{BAT Name}} & 
\colhead{\multirow{2}{*}{Counterpart Name}} & 
\colhead{R.A.\tablenotemark{b}} &
\colhead{Decl.\tablenotemark{b}} &
\colhead{\multirow{2}{*}{Source}} &
\colhead{\multirow{2}{*}{z\tablenotemark{c}}} & 
\colhead{Date} &
\colhead{Exp.\tablenotemark{d}} &
\colhead{\multirow{2}{*}{Type\tablenotemark{e}}} \\
\colhead{} &
\colhead{} &
\colhead{} &
\colhead{(deg)} &
\colhead{(deg)} &
\colhead{} &
\colhead{} &
\colhead{yyyy-mm-dd} &
\colhead{(s)} &
\colhead{} 
}
\startdata
  1 &  SWIFT J$0001.0{-}0708$ &            ${\rm 2MASXJ00004876{-}0709117}$ &           0.2032 &        $-7.1532$ &             SDSS &       0.038 &  2013-10-25 &             5400 &         Sy1.9 \\
 2 &  SWIFT J$0001.6{-}7701$ &            ${\rm 2MASXJ00014596{-}7657144}$ &           0.4420 &       $-76.9540$ &          du Pont &       0.058 &  2016-09-11 &              600 &         Sy1.5 \\
 3 &  SWIFT J$0002.5{+}0323$ &                           ${\rm NGC7811}$ &           0.6101 &           3.3519 &          du Pont &       0.025 &  2016-09-10 &              600 &         Sy1.2 \\
 4 &  SWIFT J$0003.3{+}2737$ &            ${\rm 2MASXJ00032742{+}2739173}$ &           0.8642 &          27.6547 &             SDSS &       0.040 &  2006-02-25 &             5700 &           Sy2 \\
 5 &  SWIFT J$0005.0{+}7021$ &            ${\rm 2MASXJ00040192+7019185}$ &           1.0082 &          70.3218 &    Hale 200 inch &       0.096 &  2017-08-31 &              600 &           Sy2 \\
 6 &  SWIFT J$0006.2{+}2012$ &                            ${\rm Mrk335}$ &           1.5814 &          20.2030 &    Hale 200 inch &       0.026 &  2019-08-02 &              600 &         Sy1.2 \\
 7 &  SWIFT J$0009.4{-}0037$ &           ${\rm SDSSJ000911.57{-}003654.7}$ &           2.2983 &        $-0.6152$ &             SDSS &       0.073 &  2003-05-30 &             1800 &           Sy2 \\
10 &  SWIFT J$0021.2{-}1909$ &                          ${\rm LEDA1348}$ &           5.2814 &       $-19.1682$ &          ESO-VLT &       0.096 &  2018-11-14 &      480/436/480 &         Sy1.9 \\
13 &  SWIFT J$0025.8{+}6818$ &                        ${\rm LEDA136991}$ &           6.3850 &          68.3624 &    Hale 200 inch &       0.012 &  2019-01-23 &              300 &           Sy2 \\
 14 &  SWIFT J$0026.5{-}5308$ &                        ${\rm LEDA433346}$ &           6.6695 &       $-53.1633$ &          du Pont &       0.062 &  2016-09-11 &              600 &         Sy1.5 
\enddata
\label{tab:table2}
\tablenotetext{a}{Swift-BAT 70-month hard X-ray survey ID (\url{http://swift.gsfc.nasa.gov/results/bs70mon/}).}
\tablenotetext{b}{J2000 coordinates based on WISE positions \citep{Koss_DR2_catalog}.}
\tablenotetext{c}{Input redshift used from \OIII. A full list of BASS DR2 redshifts estimated from single fits to \OIII, is provided in \citep{Koss_DR2_catalog}.}
\tablenotetext{d}{The notation for the case of ESO-VLT (X-Shooter) indicates `UVB/VIS/NIR'.}
\tablenotetext{e}{AGN classification following \citet{Winkler92}.}
\tablecomments{(This table is available in its entirety in a machine-readable form in the online journal. A portion is shown here for guidance regarding its form and content.)}
\end{deluxetable*}

Over the last five years, significant efforts have been made to implement a comprehensive and complete optical spectroscopic follow-up of the entire AGN population that was identified in the most recent Swift-BAT catalogs \citep{Baumgartner13, Oh18}. 
To this end, the BAT AGN Spectroscopic Survey Data Release 1 \citep[BASS\footnote{\url{http://www.bass-survey.com}} DR1;][]{Koss17} provided detailed measurements of the narrow and broad emission lines, stellar velocity dispersion, black hole masses ($M_{\rm BH}$), and accretion rates ($\lambda_{\rm Edd} \equiv L_{\rm bol}/L_{\rm Edd}$, where $L_{\rm Edd}$ is the Eddington luminosity: $L_{\rm Edd} \equiv 1.5 \times 10^{38}$ ($M_{\rm BH}/M_{\odot}$)) for 642 AGNs using dedicated follow-up optical and near-infrared (NIR) spectroscopic campaigns and publicly available data (the SDSS and the 6dF Galaxy Survey; \citealt{Abazajian09, Jones09, Alam15}). The DR1 dataset enabled several intriguing results on AGN physics and SMBH growth: The correlation between X-ray continuum and optical emission-lines \citep{Berney15}, a comprehensive study of the NIR for over 100 BAT AGNs \citep{Lamperti17}, and a tight relationship between Eddington ratio and \NIIHa\ emission-line ratios \citep{Oh17}.

The present study serves as part of the second data release (DR2) from BASS, and we present various spectroscopic properties of the Swift-BAT AGN that were selected from the 70-month survey catalog \citep{Baumgartner13}, such as emission-line strengths, BPT diagnostics, and AGN types, as well as reports of AGNs with double-peaked narrow lines and/or outflow signatures from \Ntotalobtained\ unique optical spectra.  An overview of the BASS DR2 survey is provided in \citet[][]{Koss_DR2_overview}, with a full description of all DR2 spectra and counterpart updates provided in \citet{Koss_DR2_catalog}. Black hole mass measurements for the BASS DR2 using broad Balmer lines \citep{Mejia_Broadlines} and velocity dispersion \citep{Koss_DR2_sigs} are also provided in separate catalogs. Finally, newly obtained DR2 NIR spectroscopy and emission line catalog (e.g. 1000\,\AA-24000\,\AA) are discussed in separate studies \citep{Ricci_DR2_NIR_Mbh,denBrok_DR2_NIR}.

The remainder of this paper is organized as follows. In Section \ref{sec:data}, we provide a brief introduction of the parent sample, telescopes, and instrumental setups used to obtain the optical spectra and a summary of the spectral reductions. In Section \ref{sec:fitting}, we describe the spectral decomposition and line fitting procedures, which include host galaxy template fitting. In Section \ref{sec:results}, we characterize the AGN demographics of the sample.  Finally, we provide a summary of our work in Section \ref{sec:summary}. Throughout this study, we assume a cosmology with $h=0.70, \Omega_{\rm M} = 0.30$, and $\Omega_{\rm \Lambda} = 0.70$.

\begin{figure} 
	\centering
	\includegraphics[width=0.99\linewidth]{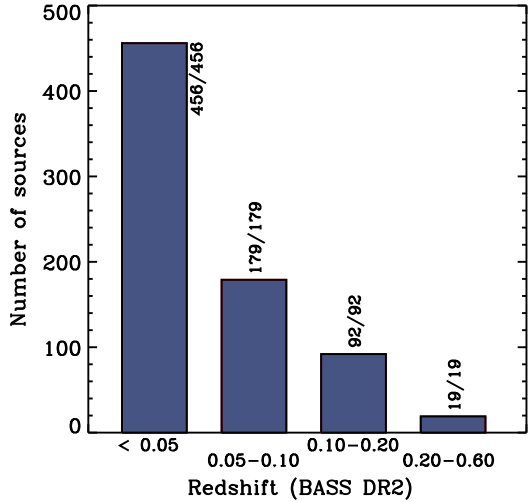} 
    \caption{Redshift distribution of the BASS sample. The number presented next to each bar indicates the number of sources in which that redshift is confirmed and the total number of sources in the given redshift range.}
    \label{fig:redshift}
\end{figure}

\section{Parent Sample and Data}
\label{sec:data}
In this section, we describe the selection of AGNs from the parent X-ray catalog and the processing and analysis of the obtained optical spectra. 

\subsection{The 70-month Swift-BAT Catalog}
\label{sec:swift}
The aim of the BASS DR2 is to provide comprehensive spectroscopic data and measurements of the AGNs identified in the 70-month Swift-BAT all-sky ultra-hard X-ray ($14-195$ keV) survey catalog\footnote{\url{http://heasarc.gsfc.nasa.gov/docs/swift/results/bs70mon/}} \citep{Baumgartner13}. 
This catalog presents 1210 objects, of which \NAGN\ sources have been identified to date as extragalactic AGNs in follow-up work (BASS DR2, \citealt{Koss_DR2_catalog, Koss_DR2_overview}). This catalog is complete across the full sky except for seven sources deep within the Galactic plane ($0<b<3^{\circ}$) with very high optical extinction values (5-43 $A_V$ mag) making optical spectroscopy impossible. Following BASS DR1 \citep{Koss17}, we limit our sample of interest to non-beamed and non-lensed AGNs by cross-matching the \NAGN\ BAT AGNs with the Roma Blazar Catalog \citep[BZCAT;][]{Massaro09} and the follow-up work by \citet{Paliya19}. This excluded beamed population includes mostly traditional continuum dominated blazars with no emission lines or host galaxy features and higher redshift ($z>0.3$) broad line quasars \citep[see][for further discussion]{Koss_DR2_overview}, which are not suitable for our optical emission line analysis. Thus, we are left with \NnonbeamedAGN\ non-beamed AGNs. 

\subsection{Optical Spectroscopic Data}
The full DR2 catalog consists of \Nspec\ optical spectra \citep{Koss_DR2_catalog}, whereas this study focuses on the single best measurement of emission-line strength. The spectra used in this study were chosen based on the wavelength range of the obtained spectra, $\sim3200$ \AA\ $-$ $\sim10000$ \AA\ in the rest-frame, which samples many prominent emission lines. We also considered S/N ratios and artifacts such as gaps and/or poor flux calibration between blue and red arms. In this study, we present $\Ntotalobtained$ optical spectra out of \NnonbeamedAGN, which is $99.6\%$ of the non-beamed AGN listed in the 70-month Swift-BAT catalog.

\begin{figure*} 
	\centering
	\includegraphics[width=1\linewidth]{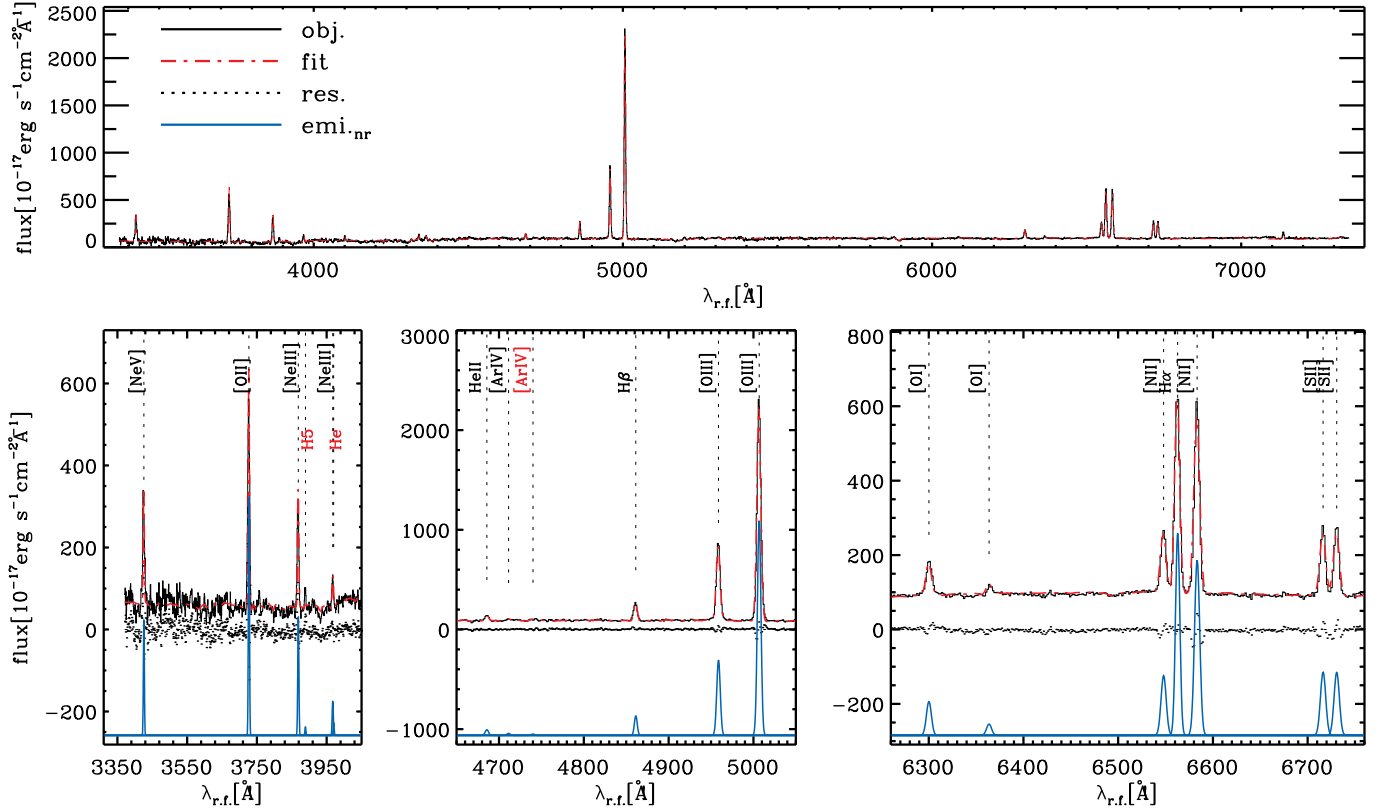} 
    \caption{Example of spectral line fitting for narrow-line source (NGC $788$). The top panel shows the full-range spectral fits. The black line represents the observed spectrum in the rest frame. The red dashed-dotted line is the best fit. The bottom panels show the spectral fitting result in detail, and they include the labels of the detected emission-lines. In the case of low A/N, smaller than 3, red labels are used. The blue Gaussians represent narrow emission-line components, which are shown with arbitrary offset for clarity. Residuals are shown using black dots.}
    \label{fig:specfits_NLR}
\end{figure*}

\begin{figure*} 
	\centering
	\includegraphics[width=1\linewidth]{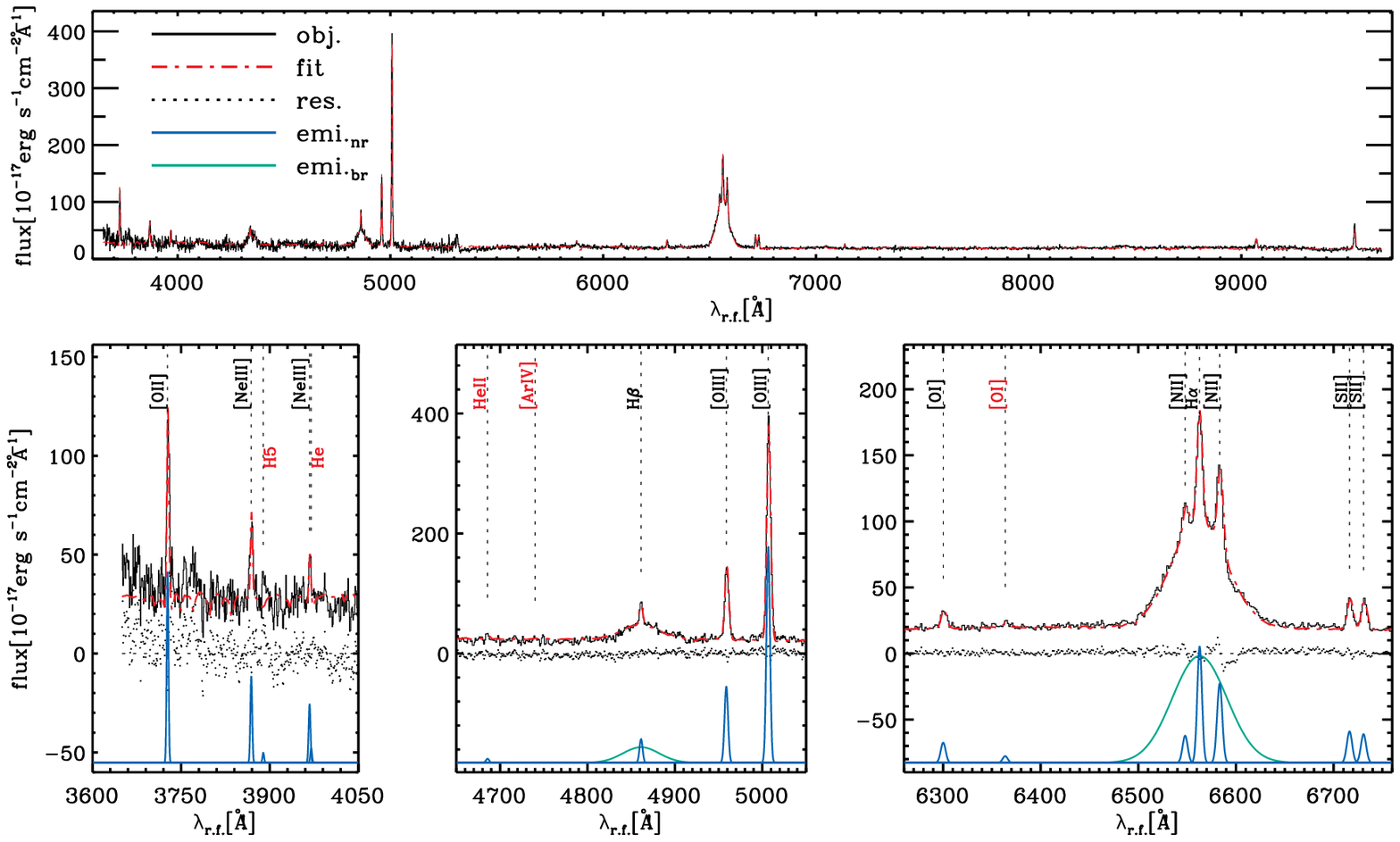} 
    \caption{Example of spectral line fitting for broad-line source (2MASX J$02593756+4417180$). The format is the same as that of Fig~\ref{fig:specfits_NLR}. The green Gaussians represent broad emission-line components.}
    \label{fig:specfits_BLR}
\end{figure*}

\subsubsection{Targeted Spectroscopic Observations}
In addition to the 225 literature spectra we used for this study (e.g., SDSS, 6dF, and BASS DR1), we also used the best available spectra from the \Nspec\ within the BASS DR2. The DR2 targeting criteria goals were to provide the largest possible sample of black hole mass measurements, including both broad line and stellar velocity dispersion measurements. The latter typically required higher resolution gratings with narrower wavelength coverage rather than the broadest possible spectral coverage (e.g. 3000--10000\,\AA), which is beneficial for emission line measurements. Here we provide a brief description of the instrumental setups for the DR2 data used in this project and subsequent data reduction procedures. Full details on the spectroscopic observations and reductions are provided in \citet{Koss_DR2_catalog}.

The largest number of optical spectra was obtained from the spectroscopic programs using the Palomar Double Spectrograph (DBSP), which is attached to the Hale 200-inch telescope ($N=\NPalomar$, 36.4\% of our AGNs). These spectra were obtained as part of a dedicated Yale program on BAT AGN (led by C. M. Urry \& M. Powell) or from observations of NuSTAR programs (led by F. Harrison \& D. Stern). Most of the observations were performed between 2012 October and 2020 November, using a 1\farcs5 slit and the 600/4000 and 316/7500 gratings. 

Another large portion of the spectra ($N=\NXshooter$, $22.7\%$) was obtained using the X-Shooter spectrograph \citep{Vernet11}, mounted on the European Southern Observatory's Very Large Telescope (ESO-VLT). 
Our extensive X-Shooter effort was executed through a series of all-weather ``filler'' programs. 
The service-mode X-Shooter observations took place between 2016 December and 2019 October, under the ESO run IDs 98.A-0635, 99.A-0403, 100.B-0672, 101.A-0765, 102.A-0433, 103.A-0521, and 104.A-0353 (led by K. Oh \& B. Trakhtenbrot). We used slits with widths of 1\farcs6, 1\farcs5, and 0\farcs9 for the UVB, VIS, and NIR arms (respectively), which provided spectral resolutions of $R=3200$, $5000$, and $5600$ for the three arms.
We employed an ABBA nodding pattern along the slit, with a nod throw of 5\farcs0. Each ABBA cycle had an exposure time of 496 s for the UVB and VIS arms and 500 s for the NIR arm. 
These single-cycle nodding patterns were repeated (1, 2, or 4 cycles) depending on the source brightness. 
The spectra were reduced using the ESO Reflex workflow (version 2.9.1, \citealt{Freudling13}), and we employed the {\tt molecfit} procedure \citep{Smette15, Kausch15} to correct for atmospheric absorption features. We also included available archival X-shooter spectra, which mainly comprised a sample of of low redshift, luminous BAT AGN ($z{<}0.01$) observed in IFU-offset mode \citep{Davies15}. The present study focuses on the optical part of the X-Shooter spectra (i.e., UVB \& VIS arms), while studies of BAT AGNs using NIR data obtained from X-Shooter observations will be provided in separate BASS publications \citep{denBrok_DR2_NIR,Ricci_DR2_NIR_Mbh}.  

We also utilized observations (led by C. Ricci) with the Boller \& Chivens (B \&\ C) spectrograph mounted on the 2.5 m Ir\'{e}n\'{e}e du Pont telescope at the Las Campanas Observatory for \NduPont\ sources in 2016 March and September. These used a 1\arcsec\ slit and a 300 lines/mm grating with a dispersion of 3.01 \AA/pixel ($3200-9084$ \AA) with a 10.4 \AA\ FWHM resolution. 

Additionally, we used observations from the Goodman spectrograph \citep{Clemens04} on the Southern Astrophysical Research (SOAR) telescope for \NSOAR\ sources between 2017 and 2020 (led by C. Ricci). A 1\farcs2 wide slit was used, providing resolutions of 5.6\,\AA\ and 3.8\,\AA\ FWHM for the 400 lines/mm and 600 lines/mm gratings, in conjunction with GG455 and GG385 blocking filters, respectively. 

Finally, five spectra were obtianed with the low-resolution imaging spectrometer (LRIS, \citealt{Oke95}) on the Keck telescope (led by D. Stern and F. Harrison). A blue grism (600 lines/mm) and red grating (400 lines/mm) were used with 1\farcs0 and 1\farcs5 slits, respectively.

\subsubsection{Archival Public Data}

The largest number of archival optical spectra (\NSDSS\ sources, $15.9\%$; see Figure~\ref{fig:fig1}) is from SDSS Data Release 15 \citep{Aguado19}. The second largest portion of the spectra ($N=\NsixDF$, $4.9\%$) is drawn from the 6dF Galaxy Survey \citep[6dFGS,][]{Jones09}. We note that the use of any measured spectral quantities originating from the optical spectra of the 6dF survey should be done with caution, owing to the lack of proper flux calibration.  

We also incorporated 66 spectra originally presented in BASS DR1, including: 
\NCTIO\ sources obtained using the $1.5$ m telescope (${\rm R}$-${\rm C}$ spectrograph) at the Cerro Tololo Inter-American Observatory (CTIO); \NSAAO\ spectra from the $1.9$ m telescope (Cassegrain spectrograph) at the South African Astronomical Observatory (SAAO), obtained as a part of the study by \citet{Ueda15}; and 33 additional spectra obtained from various telescopes and observatories (e.g., Kitt Peak National Observatory, Gemini, and Perkins, see Figure~\ref{fig:fig1}), for which the detailed instrument setups are described in a study by \citet{Koss17}.

In total, 225 optical spectra from existing archival or literature sources were used in this study.

Figure~\ref{fig:fig1} presents a summary of the \Ntotalobtained\ unique BAT AGN spectra from the BASS DR2 catalog. Instrument setups of the telescopes and spectrographs used for the BASS DR2 are summarized in Table~\ref{tab:table1}. We list the basic properties of the \Ntotalobtained\ BAT AGNs and the spectra used in this study in Table~\ref{tab:table2}.

\begin{deluxetable}{lccclcc}[ht]
\tabletypesize{\footnotesize}
\tablecaption{Ionized-gas Emission Lines 
\label{tab:emissionlines}}
\tablewidth{0pt}
\tablehead{
\colhead{No.} & 
\colhead{Species} & 
\colhead{Wavelength[\AA]} &
\colhead{} &
\colhead{No.} & 
\colhead{Species} & 
\colhead{Wavelength[\AA]}  
}
\startdata
1	&	HeII			&	3203.10 &	&	28	&	[ArIII]		&	7135.79 \\
2	&	[NeV]		&	3345.88 &	&	29	&	[OII]			&	7319.99 \\
3	&	[NeV]		&	3425.88 &	&	30	&	[OII]			&	7330.73 \\
4	&	[OII]			&	3727.03 &	&	31	&	[SXII]		&	7611.00 \\
5	&	[NeIII]		&	3868.76 &	&	32	&	[ArIII]		&	7751.06 \\	
6	&	[NeIII]		&	3967.47 &	&	33	&	HeI			&	7816.14 \\
7	&	H$\zeta$		&	3889.06 &	&	34	&	ArI			&	7868.19 \\
8	&	H$\epsilon$	&	3970.08 &	&	35	&	[FeXI]		&	7891.90 \\
9	&	H$\delta$ 		&	4101.74 &	&	36	&	HeII			&	8236.79 \\
10	&	H$\gamma$	&	4340.47 &	&	37	&	OI			&	8446.36 \\
11	&	[OIII]			&	4363.21 &	&	38	&	Pa16			&	8502.48 \\
12	&	HeII			&	4685.71 &	&	39	&	Pa15			&	8545.38 \\
13	&	[ArIV]		&	4711.26 &	&	40	&	Pa14			&	8598.39 \\
14	&	[ArIV]		&	4740.12 &	&	41	&	Pa13			&	8665.02 \\
15	&	H$\beta$		&	4861.33 &	&	42	&	Pa12			&	8750.47 \\
16	&	[OIII]			&	4958.91 &	&	43	&	[SIII]			&	8829.90 \\
17	&	[OIII]			&	5006.84 &	&	44	&	Pa11			&	8862.78 \\
18	&	[NI]			&	5197.58 &	&	45	&	[FeIII]		&	8891.91 \\
19	&	[NI]			&	5200.26 &	&	46	&	Pa10			&	9014.91 \\
20	&	HeI			&	5875.62 &	&	47	&	[SIII]			&	9068.60 \\
21	&	[OI]			&	6300.30 &	&	48	&	Pa9			&	9229.01 \\
22	&	[OI]			&	6363.78 &	&	49	&	[SIII]			&	9531.10 \\
23	&	[NII]			&	6548.05 &	&	50	&	Pa$\epsilon$	&	9545.97 \\
24	&	H$\alpha$		&	6562.82 &	&	51	&	[CI]			&	9824.13 \\
25	&	[NII]			&	6583.46 &	&	52	&	[CI]			&	9850.26 \\
26	&	[SII]			&	6716.44 &	&	53	&	[SVIII]		&	9913.00 \\
27	&	[SII]			&	6730.81 \\
\enddata
\end{deluxetable}

\begin{figure*} 
	\centering
	\includegraphics[width=1\linewidth]{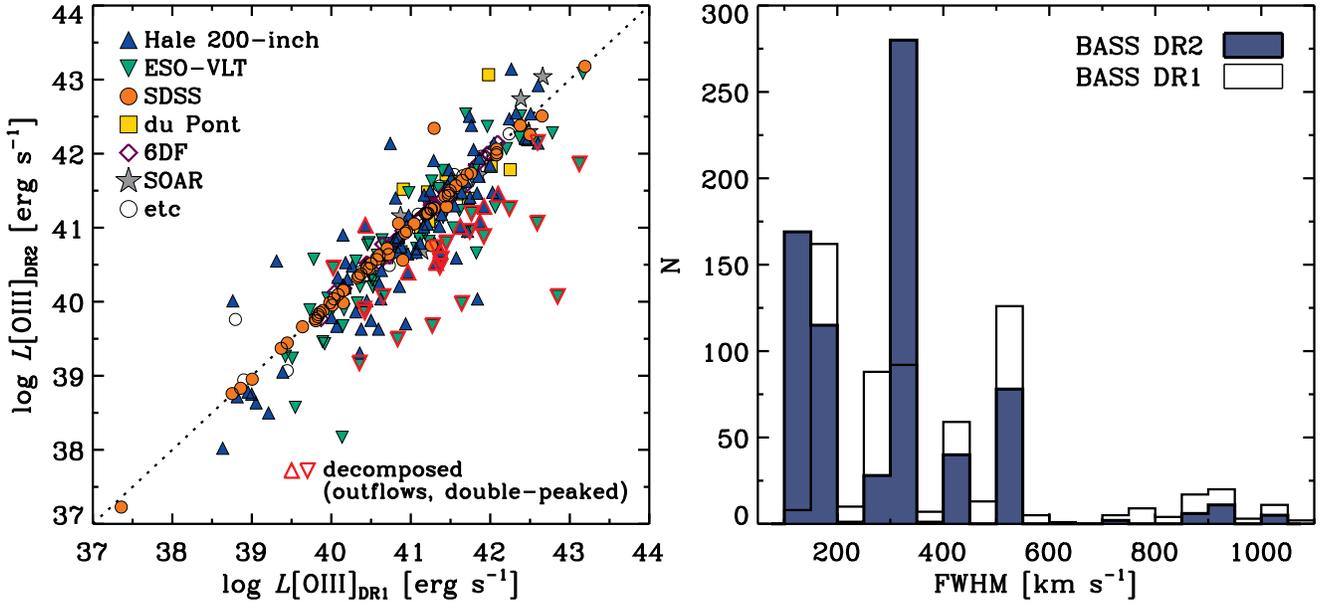} 
    \caption{Left: Comparison of \OIII\ luminosities ($L{\rm [OIII]}_{\rm DR2}$ vs. $L{\rm [OIII]}_{\rm DR1}$). Spectral sources used in DR2 are shown with different symbols and colors. Dotted line indicates the one-to-one fiducial line. Objects for which the \OIII\ line is decomposed, either to outflows or to double-peaked narrow emission lines, are shown using red thick symbols. Right: Spectral resolution (FWHM [\kms]) distribution. BASS DR2 and DR1 are shown using blue-filled and empty histograms, respectively.}
    \label{fig:OIII_DR1_DR2_comparison}
\end{figure*}

\section{Spectroscopic Measurements}
\label{sec:fitting}

Our emission line measurements and analysis of the BAT AGN optical spectra consists of three major steps, following the detailed procedures of the spectral line measurements performed by \citet{Sarzi06} and \citet{Oh11}. 
First, we de-redshifted the spectra and corrected them for Galactic foreground extinction, using the  \citet{Schlafly11} extinction maps and the \cite{Calzetti00} dust attenuation curve. 
Next, we fitted the continuum emission, and extracted the stellar kinematics, by matching the spectra with a set of stellar templates.  
We used the penalized pixel fitting method \citep[{\tt pPXF}]{Cappellari04} and employed the synthesized stellar population models \citep{Bruzual03} and the empirical stellar libraries \citep[MILES]{Sanchez06} for most of the objects, whereas the X-Shooter spectral library \citep{Chen14} was used for the X-Shooter spectra. 
More details regarding stellar velocity dispersion measurements within BASS DR2 are provided in a separate publication \cite{Koss_DR2_sigs}.
The templates were convolved and re-binned to match the spectral resolution. We masked the spectral regions that were potentially affected by nebular emission lines, skylines (5577 \AA, 6300 \AA, and 6363 \AA, rest-frame), and NaD $\lambda\lambda5890, 5896$ absorption lines in this process (Table~\ref{tab:emissionlines}). 
The masked regions cover a range of 1200 \kms, centered on the expected locations of each of the lines. 
For broad Balmer lines (\Ha, \Hb, \Hg, \Hd), a wider mask that covers the presented broad lines is used, which is typically broader than 3000 \kms.

After fitting the stellar continuum, we lifted the masks and performed a simultaneous matching of the stellar continuum and emission lines using the {\tt gandalf} code, which was developed by \citet{Sarzi06}.  {\tt gandalf} performs simultaneous emission line fitting with the galaxy template fitting used by {\tt ppxf}.
The stellar templates are well-matched to the continuum, in general, while a power-law component is adopted for 82 objects.  We note that the galaxy template fitting with {\tt ppxf} with {\tt gandalf} uses an additive and multiplicative polynomial which models out residual AGN and intrinsic dust extinction in the continuum.

We combined stellar templates with Gaussian profiles representing emission lines, using either single or multiple Gaussian templates (e.g., Balmer series), including doublets (e.g., \OIIIab\ and \NIIab). The relative strengths of some lines were set (see Table 1 in \citealt{Oh11}) based on atomic physics (\OIII, \OI, and \NII) and the gas temperature (Balmer series). We report the emission lines as observed and do not apply intrinsic galaxy extinction corrections except for in the case of the \OIII\ emission line and the usual Galactic extinction corrections .

We determined the shift and width of the Gaussian templates by employing a standard Levenberg-Marquardt optimization ({\tt MPFIT} IDL routine, \citealt{Markwardt09}). The stellar templates used in the fit were broadened by the stellar line-of-sight velocity dispersions that were derived in the previous step. Table~\ref{tab:emissionlines} presents the complete list of the emission lines included in our fits. We first atttempted to fit the spectra using only narrow components. When the fits do not well represent the given observed spectra due to underlying broad Balmer line features, we imposed additional Gaussian components with an FWHM greater than 1000 \kms. In the case of complex broad features and narrow components, we allowed a shifted line center and multiple components, if necessary. Given the scope of this study, readers are refer to \citet{Mejia_Broadlines} for parameters of broad Balmer lines (e.g., fluxes and luminosities). In order to estimate the error of the emission-line fluxes, we resampled each emission-line based on the noise 100 times and measured standard deviation. 

We provide emission-line fluxes based on the choice of a Gaussian amplitude over noise ratio (A/N) threshold of 3. In the case of less significant emission-line detections (i.e., ${\rm A/N}_{line}<3$), we list a $3\sigma$ upper limit throughout the tables (Table~\ref{tab:OIIcomplex}, Table~\ref{tab:OIIIcomplex}, Table~\ref{tab:Hacomplex}, Table~\ref{tab:SXIIcomplex}, Table~\ref{tab:SIIIcomplex}). An example of a spectral line fit is shown in Figure~\ref{fig:fits}. The spectral fits of all BAT AGNs that were analyzed in this study are available on the BASS Website.

\begin{figure} 
	\centering
	\includegraphics[width=0.99\linewidth]{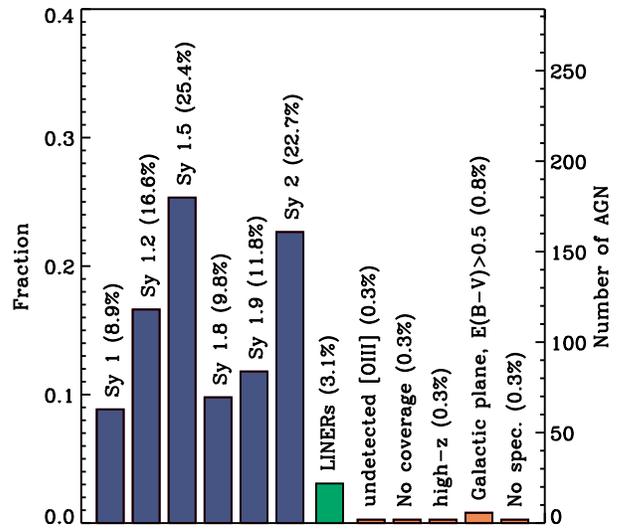} 
    \caption{AGN sub-classes (blue), LINERs (green) and unclassifiable cases (orange). We classified $98\%$ of BAT AGNs (732/746) into the subtypes. `Undetected [OIII]' refers to objects that lack the \OIII\ emission line. `No coverage' indicates objects for which either \OIII\ falls into a spectral gap between the blue and red sides of the detector or \Ha\ is lacking in the spectral complex, owing to a limited spectral coverage of the obtained optical spectra. `High-z' represents objects that are not applicable for classification using the obtained spectral coverage, owing to their high redshift nature. `Galactic plane' indicates objects with a poor fitting quality at E(B$-$V) $>0.5$.}
    \label{fig:AGNtypes}
\end{figure}

\begin{figure} 
	\centering
	\includegraphics[width=0.99\linewidth]{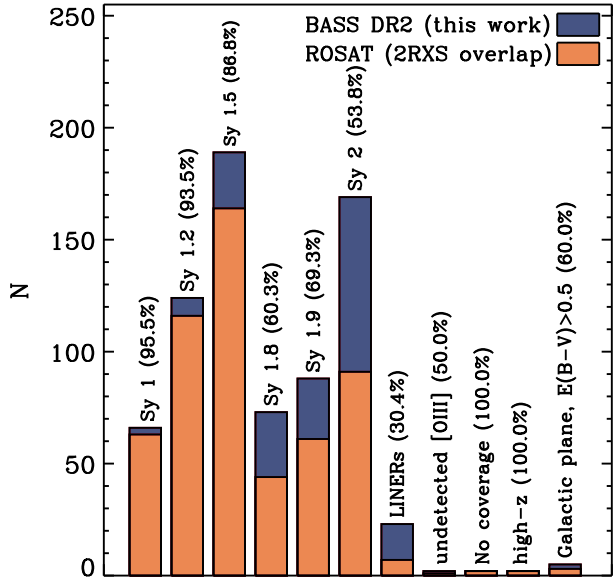} 
    \caption{Subtypes of BAT AGNs (blue) in common with the second ROSAT all-sky survey (2RXS, orange) source catalog. In total 554 sources are found in both catalogues out of 743 BAT AGNs discussed in this study.}
    \label{fig:rosat}
\end{figure}

\section{Results}
\label{sec:results}

\subsection{Redshift Distribution}
We used the 105-month survey redshifts as input for spectral line fits, and we manually adjusted them when necessary. For the objects with unknown redshift, we estimated the redshift using either the peak of \OIII\ or narrow \Ha\ emission lines considering the presence of \OIII\ outflows. A full list of redshifts estimated from single emission line fits to \OIII, and errors is provided in \citep{Koss_DR2_catalog}, and this should be used for NLR emission offset studies.

Figure~\ref{fig:redshift} presents the redshift distribution of the BAT AGNs at different redshift intervals. As with DR1, the majority of BAT AGNs are nearby objects detected at $z < 0.2$ ($\sim 97\%$). We achieve $100\%$ completeness in redshift determination for the non-beamed AGNs from the 70-month BAT AGN catalog (3 sources are not included but have redshifts due to being deep within the Galactic plane or have foreground stellar contamination). The median $z$ of the BAT AGNs presented in this study is $0.038$. Note that we determine new redshifts for 82 BAT AGNs not listed in the NASA/IPAC Extragalactic Database (See \citealt{Koss_DR2_catalog} for full details).

\begin{figure*} 
	\centering
	\includegraphics[width=0.7\linewidth]{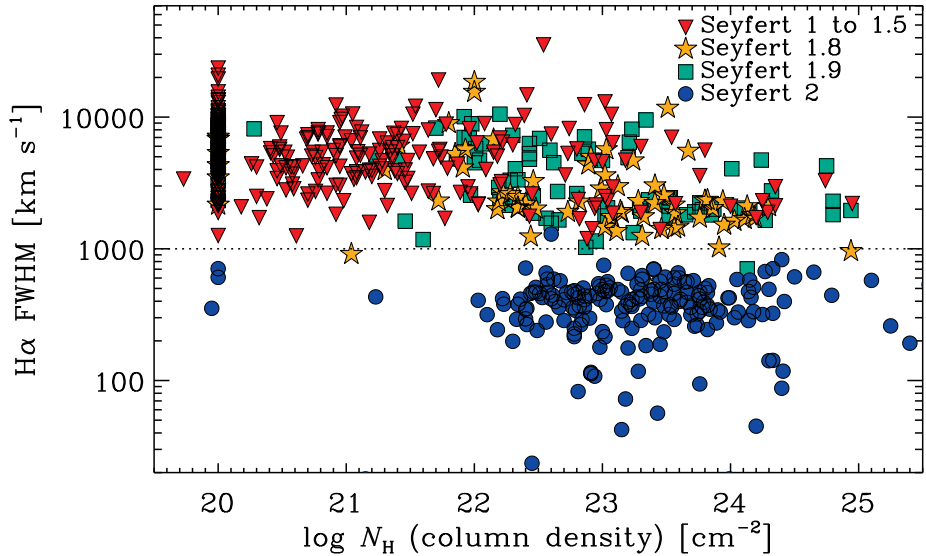} 
    \caption{FWHM of \Ha\ as a function of hydrogen column density. There are 199 examples of Sy1-1.5, 7 of Sy1.8, and 9 of Sy1.9 that exhibit a column density with a lower limit ($N_{\rm H}=10^{20}~{\rm cm}^{-2}$). The colors represent AGN subtypes according to the Winkler classification \citep{Winkler92}.}
    \label{fig:broad_Ha_FWHM_vs_NH}
\end{figure*}

\subsection{Comparison of \OIII\ with the DR1}

We use the new BASS DR2 data and spectral measurements to compare the luminosity of the \OIII\ line (\LOIII\ hereafter) with that measured as part of DR1, as shown in Figure~\ref{fig:OIII_DR1_DR2_comparison}. Overall, the DR2 \LOIII\ measurements appear to be highly consistent with the DR1 ones. This is particularly evident for the SDSS and 6dF spectra that we reused by applying a different spectral line fitting procedure compared with DR1. Note that, unlike DR1, we performed a full-range spectral fitting to measure emission-line strengths considering underlying stellar components in this study. The higher-resolution spectra that were obtained as part of DR2 (e.g., ESO-VLT X-Shooter) are the main source of scatter. The asymmetric scatter towards lower \LOIII\ in DR2 is caused by more reliable emission-line decomposition, which can better account for outflow components (see detailed study of \citealt{Rojas20}) or double-peaked narrow emission-line features (red thick symbols in the left panel of Fig.~\ref{fig:OIII_DR1_DR2_comparison}); this is a direct consequence of the high quality of the spectra. 
These optical spectra with higher resolution than those in DR1 are distinctively exhibited in the right panel of Fig.~\ref{fig:OIII_DR1_DR2_comparison}.

\begin{figure*} 
	\centering
	\includegraphics[width=0.8\linewidth]{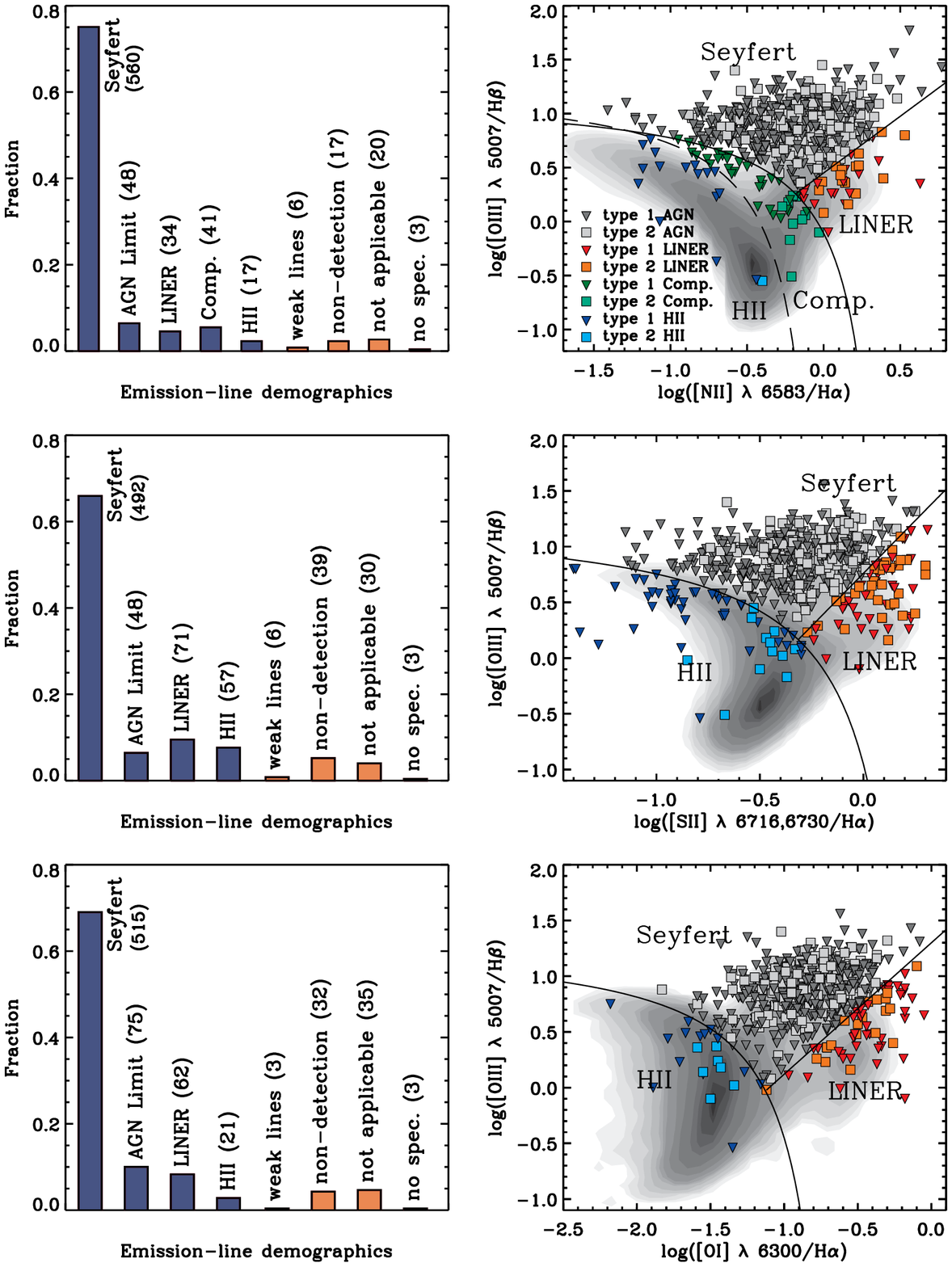} 
    \caption{Emission-line classification of the BAT AGNs using line diagnostics diagrams \citep{Baldwin81, Kauffmann03, Kewley01, Kewley06, Schawinski07}. Left panels: histograms for the entire sample. `AGN limit' refers to objects that are located either in the Seyfert or LINER regions, which have a $3\sigma$ upper limit in any of the used emission lines. The remaining categories, shown using orange histograms, have features that prevented measurements. `Weak lines' includes objects where the Gaussian amplitude to noise ratios (A/N) of the detected emission lines do not surpass the threshold (${\rm A/N}_{line}<3$). `Non-detection' refers to objects not detected in both of the needed emission-lines to measure a ratio. `Not applicable' represents objects for which the emission-line strengths are less reliable. This may be owing to either the high E(B$-$V) at the Galactic plane ($>0.5$) or the lack of spectral coverage, owing to high-z nature or instrumental limitations. Objects with a poor spectral fit in any of the used emission lines and the two objects without the \OIII\ and \Ha\ spectral complex are also included in this category. `No spec.' indicates three missing spectra. The number of sources in each class is written in parenthesis. Right panels: line diagnostic diagrams for sources with sufficient measurable emission lines to be classified using line diagnostic diagrams. Narrow-line objects are shown with squares and broad-line objects with triangles. The grey filled contours are drawn from the OSSY catalog and the follow-up broad-line AGN study that shows the samples of SDSS emission-line galaxies at $z<0.2$ \citep{Oh11, Oh15}. Note that only narrow lines are used to produce the diagnostics diagrams.}
    \label{fig:bpt1}
\end{figure*}

\begin{figure*} 
	\centering
	\includegraphics[width=0.84\linewidth]{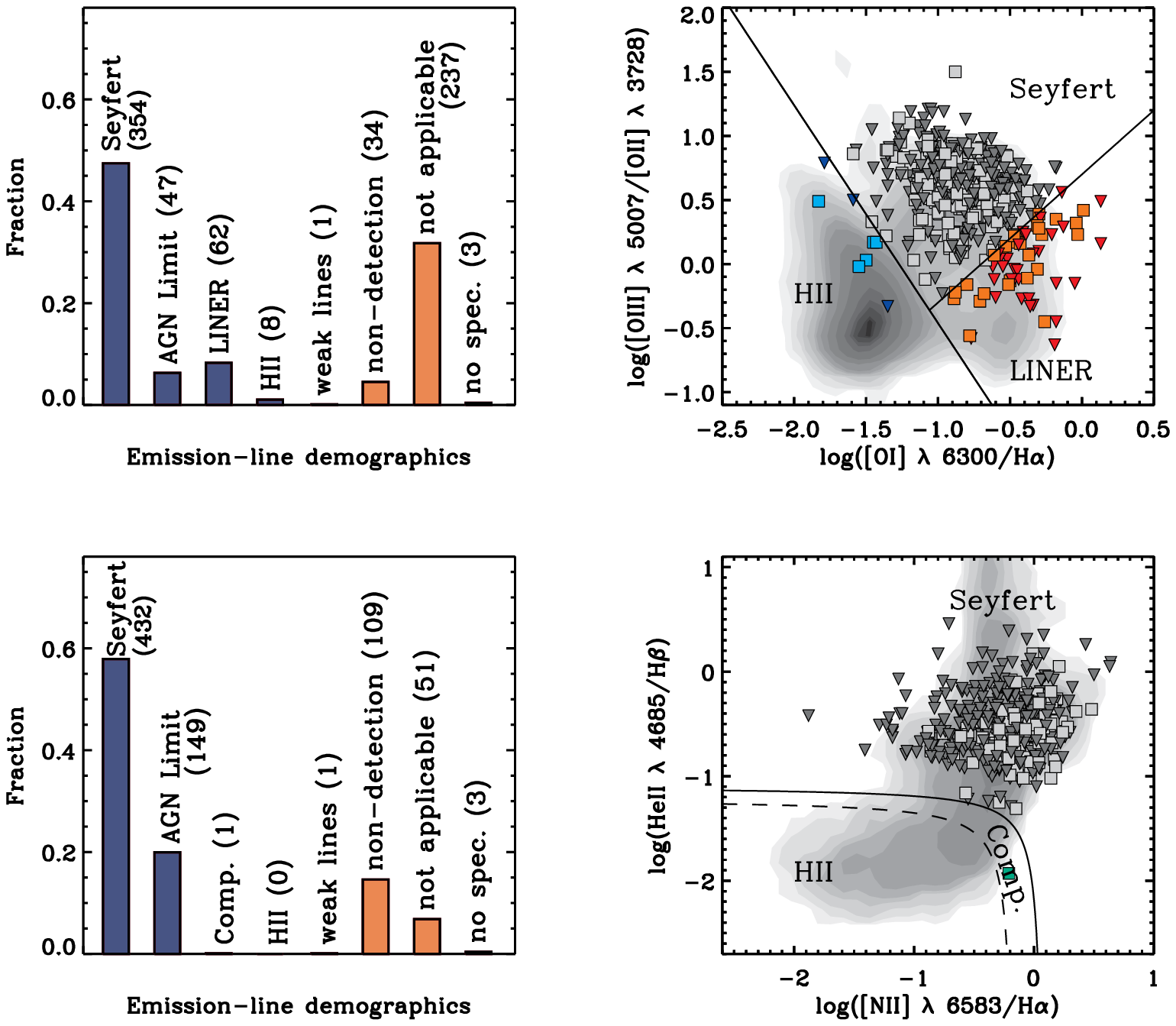} 
    \caption{Emission-line classification of the BAT AGNs using line diagnostics diagrams \citep{Kewley06, Shirazi12}. The same scheme as the previous line diagnostic figure is used.}
    \label{fig:bpt2}
\end{figure*}

\subsection{AGN Type Classification}
\label{sec:AGN_type_classification}
We classified the subtypes of AGNs based on the presence of broad-line emission and flux ratios between \Hb\ and the \OIII\ emission lines, following studies by \citet{Osterbrock81} and \citet{Winkler92}. The Seyfert 2 classification refers to a source without broad-line emissions. A source that lacks broad lines in \Hb\ yet exhibits a broad signature in \Ha\ is classified as Seyfert 1.9. The remaining Seyfert subtypes (1, 1.2, 1.5, and 1.8) were determined using the total flux of \Hb\ and \OIII\ \citep{Winkler92}.

Figure~\ref{fig:AGNtypes} presents BAT AGN subtypes according to fraction and number (blue), as well as objects for which classification is not applicable (orange). `Not applicable' sources may occur owing to various reasons, such as a lack of \OIII\ emission lines used in the classification, limited spectral coverage (i.e., spectral setups, high-$z$ nature), and high E(B-V) in the Galactic plane. Compared with DR1, the overall number of `not applicable' sources markedly decreased from 103 to 14, representing only 1.9 $\%$ of all non-beamed BAT AGNs ($14/746$).

In Fig~\ref{fig:rosat}, we show the subtypes of AGNs in common ($554/743$, $75$\%) with the second ROSAT all-sky survey (2RXS) source catalog \citep{Boller16}. Due to its capability in detecting unobscured AGN in the soft X-ray band ($0.1-2.4$ keV, \citealt{Truemper82}), the majority of the BAT AGNs are found in Seyfert type featuring broad-lines in their optical spectra. The overlap with ROSAT gradually declines with Sy1.9 and Sy2, and finally LINERs, consistent with their higher average column densities that absorbs the soft X-rays leading to their non-detection in ROSAT. The frequency of Sy1 to Sy2 in the sample of ROSAT is known to about 11:1 \citep{Kollatschny08}. The nature of these ROSAT X-ray detected AGNs, including such a dominant incidence of broad-line sources, have been reported by numerous studies (e.g., \citealt{Pietsch98, Zimmermann01, Kollatschny08}). A key point to keep in mind, however, is that while the detection overlap is relatively good, the X-ray fluxes and related properties derived from the 2RXS tend to be systematically low by up to 2 dex for the most extreme obscured AGN. A more comprehensive and complete comparison between the 2RXS and the BAT AGNs is available in \citet{Oh18}. 

\begin{deluxetable*}{lllllll}
\tablecaption{Emission Line Classification}
\tablewidth{0pt}
\tablecolumns{7}
\tablehead{
\colhead{ID\tablenotemark{a}} & 
\colhead{Counterpart Name} & 
\colhead{\NII/\Ha}  &
\colhead{\SII/\Ha} &
\colhead{\OI/\Ha} &
\colhead{\HeII} &
\colhead{\OIII/\OIIa} 
}
\startdata
1 &         2MASXJ00004876-0709117 &              Seyfert &              Seyfert &              Seyfert &              Seyfert &              Seyfert \\
2 &         2MASXJ00014596-7657144 &              Seyfert &              Seyfert &              Seyfert &              Seyfert &              Seyfert \\
3 &                        NGC7811 &              Seyfert &              Seyfert &              Seyfert &           Weak lines\tablenotemark{b} &           Weak lines \\
4 &         2MASXJ00032742+2739173 &              Seyfert &              Seyfert &              Seyfert &              Seyfert &              Seyfert \\
5 &         2MASXJ00040192+7019185 &              Seyfert &              Seyfert &              Seyfert &            AGN Limit\tablenotemark{c} &              Seyfert \\
6 &                         Mrk335 &           Weak lines &                  HII &              Seyfert &           Weak lines &            AGN Limit \\
7 &        SDSSJ000911.57-003654.7 &              Seyfert &              Seyfert &              Seyfert &              Seyfert &              Seyfert \\
10 &                       LEDA1348 &              Seyfert &              Seyfert &              Seyfert &              Seyfert &            AGN Limit \\
13 &                     LEDA136991 &            AGN Limit &            AGN Limit &            AGN Limit &           Weak lines &           Weak lines \\
14 &                     LEDA433346 &              Seyfert &              Seyfert &              Seyfert &              Seyfert &              Seyfert
\enddata
\label{tab:BPT}
\tablenotetext{a}{Swift-BAT 70-month hard X-ray survey ID (\url{http://swift.gsfc.nasa.gov/results/bs70mon/}).}
\tablenotetext{b}{`Weak lines' refers to objects lacking sufficiently strong emission-line strengths with ${\rm A/N}<3$ to be placed on the diagnostic diagrams.}
\tablenotetext{c}{`AGN Limit' refers to objects classified as either Seyfert or LINER having a $3\sigma$ upper limit in the used emission lines.}
\tablecomments{(This table is available in its entirety in a machine-readable form in the online journal. A portion is shown here for guidance regarding its form and content.)}
\end{deluxetable*}

We also show the FWHM of the \Ha\ emission line as a function of the hydrogen column density that is derived from the X-rays \citep[$N_{\rm H}$]{Ricci17} in Figure~\ref{fig:broad_Ha_FWHM_vs_NH}. Approximately half of the Seyfert $1$-$1.5$ AGNs ($199/379$) have upper limits on $N_{\rm H}=10^{20}$ ${\rm cm}^{-2}$, supporting their unobscured nature. Most unobscured Seyfert $1$-$1.5$-type AGNs were observed with $N_{\rm H}\le10^{22}$ ${\rm cm}^{-2}$ ($85\%, 323/379$), whereas Seyfert $1.9$-$2$-types predominantly have $N_{\rm H}\ge10^{22}$ ${\rm cm}^{-2}$ ($91\%, 252/278$). Only $8\%$ ($60/\Ntotalobtained$) of BAT AGNs present broad-line signatures in the optical band with $N_{\rm H}\ge10^{22}$ ${\rm cm}^{-2}$. A recent study by \citet{Ogawa21} explained the presence of these anomalous sub-populations using dust-free gas inside the torus region. Moreover, most Seyfert $1.8$-$1.9$-types ($78\%$, $125/161$) exhibited $N_{\rm H}\ge10^{22}$ ${\rm cm}^{-2}$, confirming the obscured nature of these AGNs; this is in agreement with the lack of or very weak broad \Hb\ emission lines, as in \citet{Burtscher16}. In conclusion, Figure~\ref{fig:broad_Ha_FWHM_vs_NH} illustrates the dominant conformity of the AGN type classification based on optical and X-ray spectral analysis, with a few still-debated cases in the context of the AGN standard unification model \citep{Lusso13, Merloni14, RicciF17}.

\subsection{Emission-line Classification}
\label{sec:emission_classification}
We investigated narrow emission-line diagnostics using BPT diagrams \citep{Baldwin81}. In all panels presented in \cref{fig:bpt1,fig:bpt2}, we used ${\rm A/N}_{line} >3$ to determine the significance of the line strengths. Figure~\ref{fig:bpt1} shows three diagnostic diagrams (\OIII/\Hb\ versus \NII/\Ha, \SII/\Ha, and \OI/\Ha) that employ the demarcation lines used by \citet{Kauffmann03}, \citet{Kewley01, Kewley06}, and \citet{Schawinski07}. Most of the BAT AGNs presented in this study lie in the Seyfert region of the \NII\ diagram ($75.4\%$, $560/\Ntotalobtained$). The second largest subgroup comprises objects with an upper limit in any of the used emission lines ($6.5\%$, $48/\Ntotalobtained$), which suggests either Seyfert or LINER classification. The \NII/\Ha\ versus \OIII/\Hb\ diagnostic diagram works well in general; it leaves only $2.3\%$ $(17/\Ntotalobtained)$ of the ultra-hard X-ray selected AGN in the HII region. As a comparison, the SDSS emission-line galaxy samples at $z<0.2$ (the OSSY catalog\footnote{\url{http://gem.yonsei.ac.kr/ossy}}, \citealt{Oh11,Oh15}) are shown together with the BAT AGNs in \cref{fig:bpt1,fig:bpt2}. Note that objects which exhibit weak emission lines (i.e., ${\rm A/N}_{line}<3$) or lack emission lines, owing to insufficient spectral coverage, are excluded from the analysis; this applies to less than $10\%$ of the objects in all three diagnostic diagrams.

\begin{figure*} 
	\centering
	\includegraphics[width=1\linewidth]{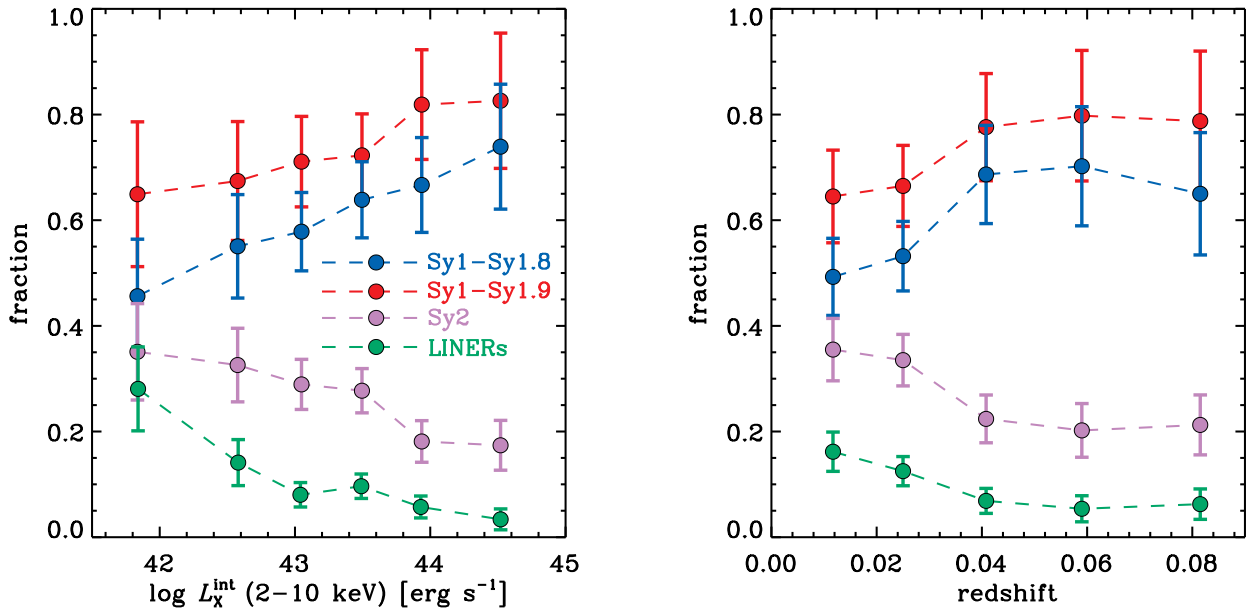} 
    \caption{Fraction of sub-classes of Seyferts and LINERs as a function of $2$-$10$ keV intrinsic luminosity (left panel) and redshift (right panel). Blue filled dots (median at each bin) and a dashed line present Seyfert $1$-$1.8$, which display broad \Hb, whereas AGNs with broad \Ha\ are shown in red. Light purple dots and a dashed line represent Seyfert $2$ narrow-line AGN. Note that the fraction of LINERs selected from the \SII/\Ha\ diagnostic diagram is shown with green.}
    \label{fig:fraction_L0210}
\end{figure*}

\begin{figure*} 
	\centering
	\includegraphics[width=1\linewidth]{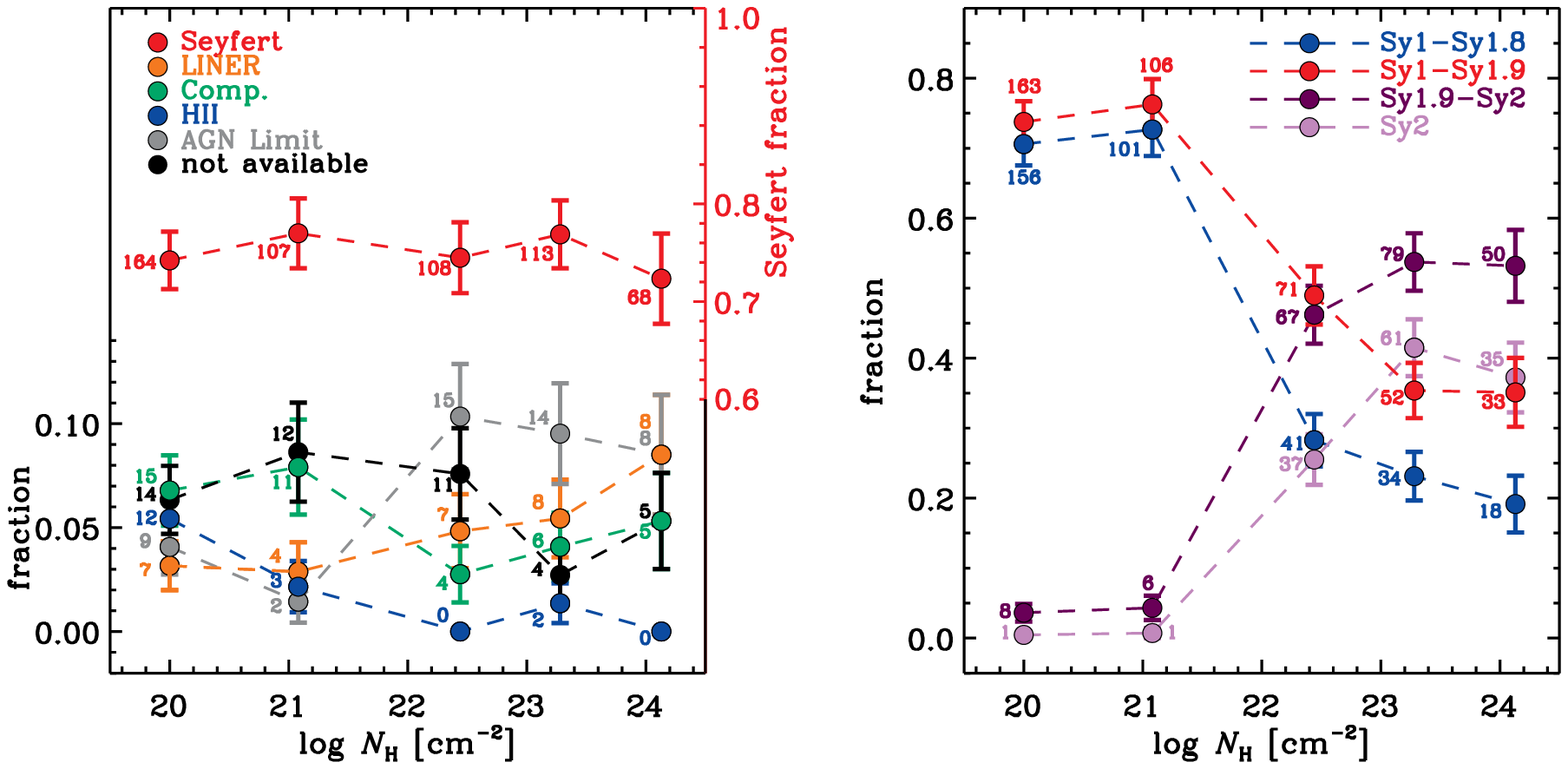} 
    \caption{AGN-type fraction as a function of column density. The \OIIIHb\ vs. \NIIHa\ diagram is used for classification. Left: The left ordinate is for objects classified by emission-line ratios as LINER (orange), composite (green), HII (blue), AGN with an upper-limit (grey), and objects that cannot be diagnosed, such as `Galactic plane', `high-z', `undetected \OIII', `no spec.', and objects exhibiting weak emission-lines (${\rm A/N}_{\rm line}<3$) (black). The right ordinate is for the Seyfert class (red). The total number of objects presented in each bin is 221, 139, 145, 147, and 94, from low to high column density. The number of objects in each class at a given column density bin is shown in the figure next to its fraction. Right: Seyfert AGNs shown in the left panel are further classified into subtypes. }
    \label{fig:nH_AGN_type_fraction}
\end{figure*}

The \SII/\Ha\ diagnostic does not appear different from the \NII/\Ha\ diagram; it displays a similar distribution of BAT AGNs with a slightly lower fraction of the Seyfert class ($66.2\%$, $492/\Ntotalobtained$). This is explained by weaker \SII\ line strengths and/or a higher fraction of HII than that of the \NII/\Ha\ diagnostic. This is also true in the case of the \OI/\Ha\ diagnostic, which yields a $69.3\%$ ($515/\Ntotalobtained$) Seyfert fraction. 

Two additional diagnostic diagrams are shown in Figure~\ref{fig:bpt2}: \OIII/\OIIa\ versus \OI/\Ha, and \HeII/\Hb\ versus \NII/\Ha. These diagnostics are less efficient in classifying Seyfert AGNs compared with the commonly used methods shown in Figure~\ref{fig:bpt1}. 

The primary reason for the high fraction of `not applicable' is a poor spectral quality at the blue end of the obtained spectra, which results in poor fitting quality and insignificant line strengths (A/N$<3$). Similar to \OIIa, \HeII\ is difficult to detect. We identified $270$ cases of low A/N ($<3$) in \HeII. These complications naturally lead to a notably high fraction of `non-detection', `AGN limit', and `not applicable' cases in these diagnostics. Table~\ref{tab:BPT} summarizes emission-line classifications.

\subsection{AGN type fraction}
Subsequent to classifying AGN types as presented in Sections \ref{sec:AGN_type_classification} and \ref{sec:emission_classification}, we present the AGN-type fraction versus $2$-$10$ keV intrinsic luminosity in Figure~\ref{fig:fraction_L0210}. The intrinsic luminosity between $2$ and $10$ keV was used, which was determined through detailed X-ray spectral fitting, as described by \citet{Ricci17}. The broad-line AGN fraction (type 1 AGN fraction) as a function of the $2$-$10$ keV intrinsic luminosity, which is a proxy of AGN bolometric luminosity, was further examined for the presence of broad Balmer lines (\Hb, blue filled dots; \Ha, red filled dots). Both cases clearly show a general increase in the broad-line AGN fraction with increasing $2$-$10$ keV intrinsic luminosity, which is consistent with previous literature \citep{Merloni14}. In contrast, the fraction of LINERs implied from the \OIII/\Hb\ versus \SII/\Ha\ diagnostic diagram decreased with increasing X-ray luminosity. A similar trend was observed for AGNs at $z<0.1$, which is in agreement with earlier studies \citep{Lu10, Oh15}. It should be noted that the decreasing fraction of LINERs in the higher redshift regime is owing to the BAT sensitivity limit.

AGN types are addressed in further detail in Figure~\ref{fig:nH_AGN_type_fraction} as a function of the hydrogen column density, which is determined by X-ray spectral analysis \citep{Ricci17}. Most objects ($>70\%$) are classified as Seyfert, as illustrated in Figure~\ref{fig:bpt1}, which is shown in red in the left panel of Figure~\ref{fig:nH_AGN_type_fraction}. The abundance of Seyfert AGNs is approximately constant over a wide range of column densities, ranging from the Compton thin to the Compton thick regimes; however, the other AGN classifications are infrequent, at less than $10\%$ of the objects. Despite the low fraction, we observed that the `AGN limit' sources increase at $N_{\rm H}>10^{22}$~$\rm{cm^{-2}}$. This can be interpreted as the result of obscuration affecting observed strengths of line intensities. The right panel of Figure~\ref{fig:nH_AGN_type_fraction} depicts the Seyfert subtypes. A clear dichotomy is displayed between the Seyfert subtypes and X-ray obscuration  (\citealt{RicciF17}, and references therein), in agreement with the classical AGN unified model. We do find some Sy$1.9$ sources and one Sy$2$ AGN (dark purple and light purple symbols) with little absorption ($N_{\rm H}<10^{22}$~${\rm cm^{-2}}$) (e.g., \citealt{Ptak96, Bassani99, Pappa01, Panessa02}), which may imply the possible disappearance of the broad-line region at a low accretion state \citep{Nicastro03, Elitzur09}. We note that \citet{Bianchi19} reported the presence of the weak broad \Ha\ line from NGC 3147, questioning the existence of true Sy2 AGN (i.e., unobscured X-ray Sy2 AGNs without a broad-line region). Alternatively, the difference in X-ray and optical obscuration classification may be the result of changing optical type AGN \citep[e.g.,][]{Collin-Souffrin:1973:343,Shappee:2014:48} where the non-simultaneous measurements of X-ray and optical spectroscopy are tracing intrinsic variability.

\begin{figure} 
	\centering
	\includegraphics[width=0.99\linewidth]{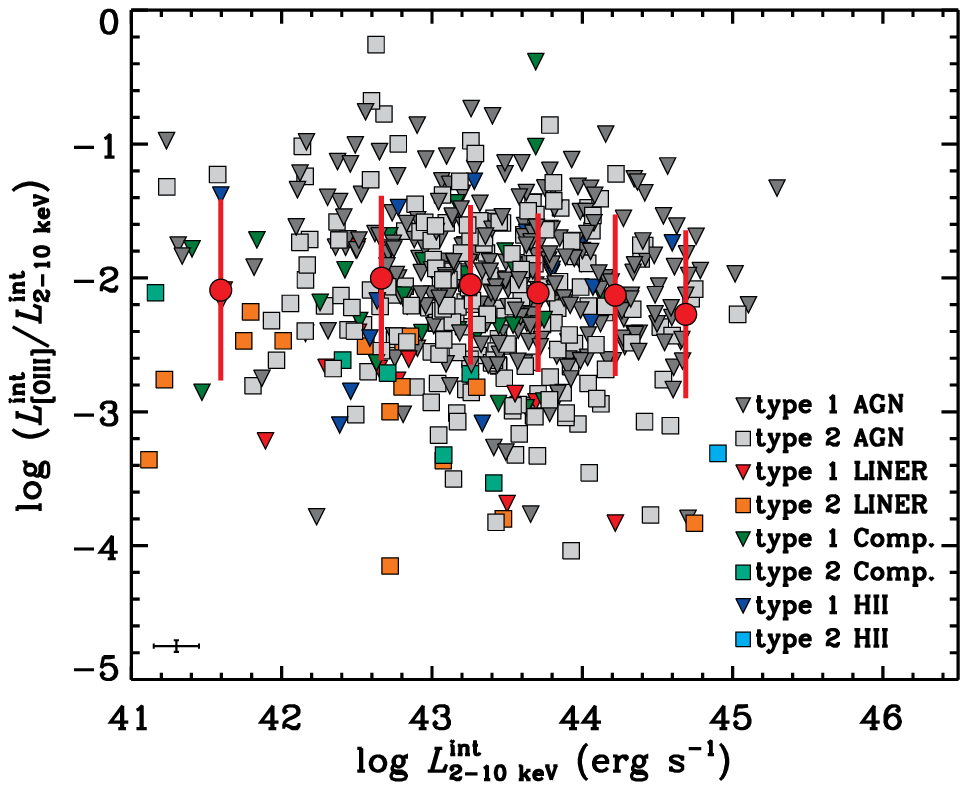} 
    \caption{$L^{\rm int}_{\rm [OIII]}/L^{\rm int}_{\rm 2-10 keV}$ ratio as a function of $2$-$10$ keV intrinsic luminosity. The red filled dots indicate average values, and the red bars show 1$\sigma$ deviations in the bins. This figure includes BAT AGNs that are classified into either one of the categories (Seyfert, LINER, composition, and HII) from the \OIII/\Hb\ vs. \NII/\Ha\ diagnostic diagram. Typical uncertainties are shown in the bottom-left corner. The $2-10$ keV intrinsic luminosity bins are of equal width. The symbols and colors are the same as that of Figure~\ref{fig:bpt1}.}
    \label{fig:L0210_LOIIIoverL0210}
\end{figure}

\subsection{$L^{\rm int}_{\rm [OIII]}/L^{\rm int}_{\rm 2-10~keV}$ ratio versus X-ray luminosity}
Figure~\ref{fig:L0210_LOIIIoverL0210} shows the $L^{\rm int}_{\rm [OIII]}/L^{\rm int}_{\rm 2-10~keV}$ ratio as a function of $2$-$10$ keV intrinsic luminosity. The average values and 1$\sigma$ deviation in the bins are presented as red filled dots and bars, respectively. The general trend of the $L^{\rm int}_{\rm [OIII]}/L^{\rm int}_{\rm 2-10 keV}$ is consistent over $L^{\rm int}_{\rm 2-10 keV}$ with a slight decrease. However, this decrease in the average values from low to high luminosity was not statistically significant. The average values of the $L^{\rm int}_{\rm [OIII]}/L^{\rm int}_{\rm 2-10 keV}$ at the lowest and the highest quartile of $L^{\rm int}_{\rm 2-10~keV}$ are $-2.02$ and $-2.19$, respectively.  Due to the different slit widths and various redshifts the NLR measurements may extend between 200 pc and the size of the entire galaxy.  In addition, the NLR size may vary depending on the power of the AGN (e.g. \citealt{Hainline13}).  While important these corrections have not been found to be more than 0.1-0.2 dex in the nearest systems within BASS (e.g. \citealt{Ueda15, Berney15}).  Further efforts, such as ongoing efforts with large FOV IFUs such as VLT/MUSE are neccesary to fully study these issues.

\begin{figure*} 
	\centering
	\includegraphics[width=0.88\linewidth]{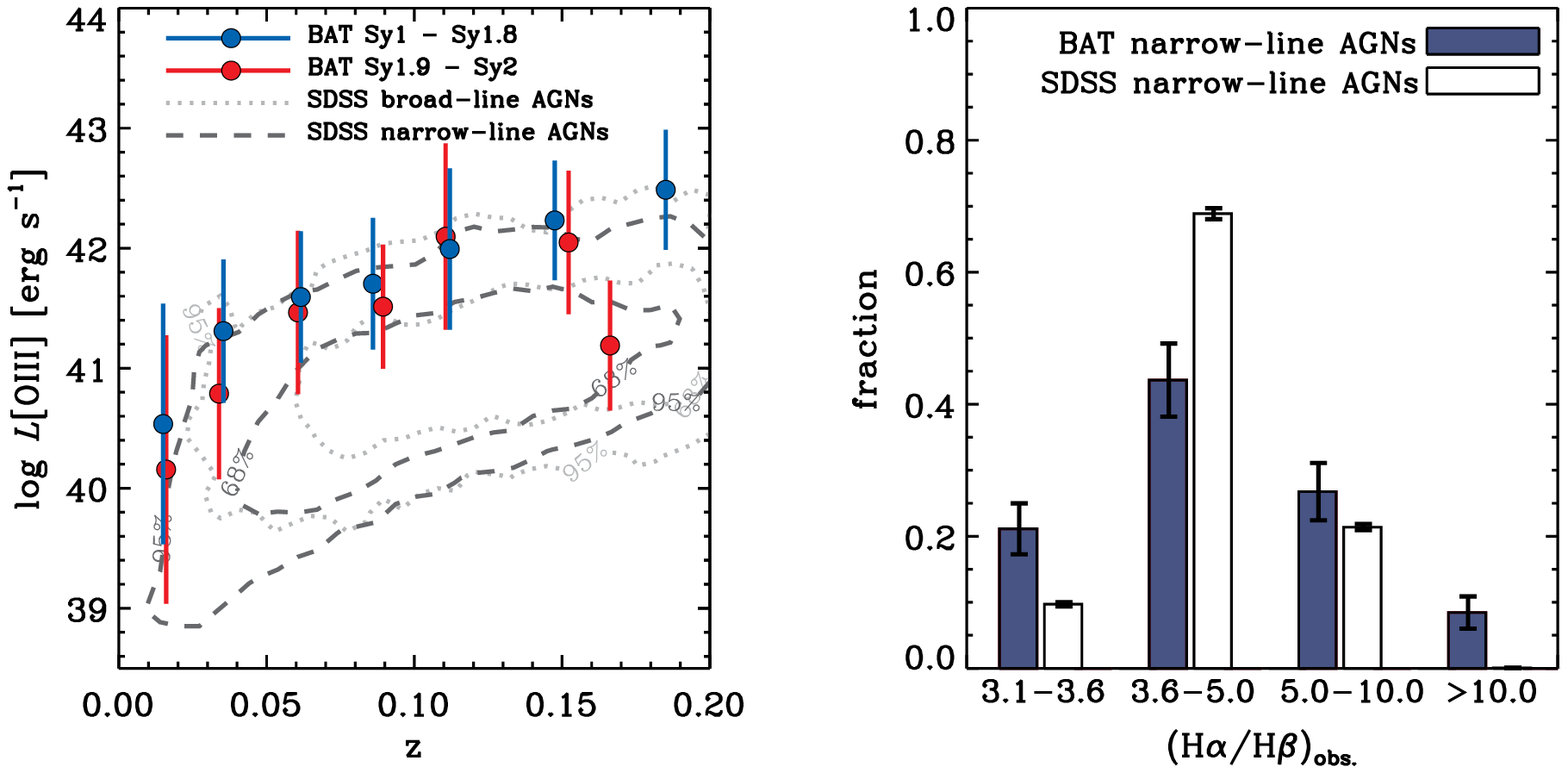} 
    \caption{Comparison of BAT AGNs with optically selected AGNs in the SDSS from the OSSY catalog \citep{Oh11}. Left panel: \OIII\ luminosity vs. redshift. Median and 1$\sigma$ distribution for BAT AGNs are shown in blue (Sy1$-$Sy1.8) and red (Sy1.9 and Sy2), respectively. The SDSS AGNs are displayed in light grey dotted lines (broad-line AGNs) and thick dark grey dashed lines (narrow-line AGNs). The contours represent 68$\%$ and 95$\%$ distributions. The BAT broad-line AGN and narrow-line AGNs exhibit higher \OIII\ luminosities than the SDSS AGNs. Right panel: Balmer decrement compared to the SDSS narrow-line AGNs. BAT narrow-line AGNs are dustier than the SDSS narrow-line AGNs, which exhibit a higher fraction of $({\rm H}\alpha/{\rm H}\beta)_{\rm obs.}$ greater than 5.}
    \label{fig:LOIII_z_Balmer_decrement}
\end{figure*}

\begin{figure*} 
	\centering
	\includegraphics[width=0.88\linewidth]{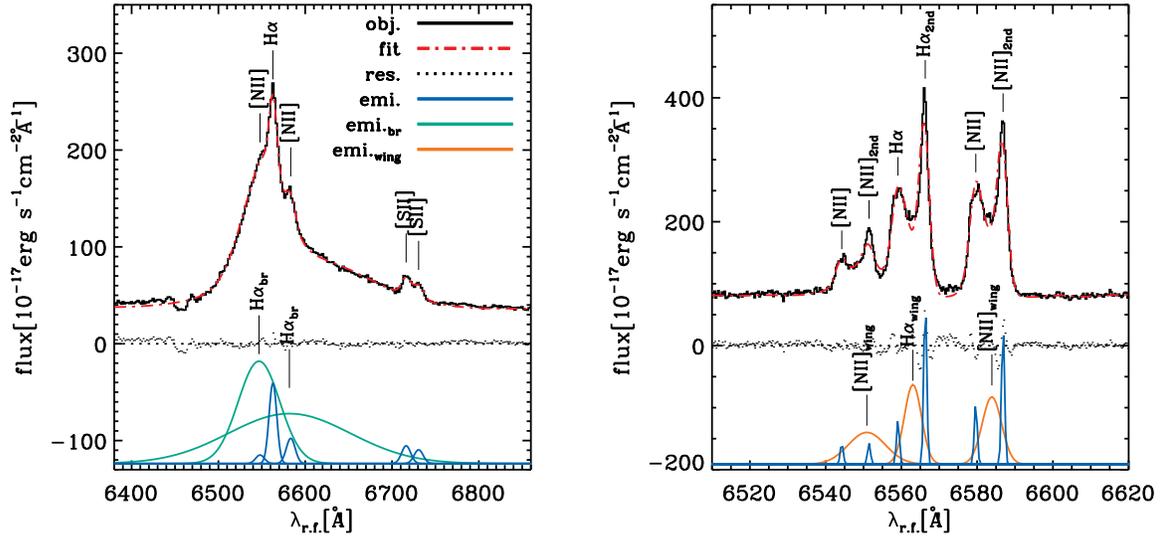} 
    \caption{Examples of AGNs featuring an asymmetric broad-lines and double-peaked narrow emission-lines. Left panel: \Ha\ spectral complex of RBS 273, which presents the asymmetric broad-line components. The color codes are the same as that of Figure~\ref{fig:fits}. Right panel: \Ha\ spectral complex of IC 4709. Double-peaked narrow emission-lines (blue) with underlying wing components (orange) are distinctively decomposed. The measured velocity offset between the narrow emission-lines is 328 \kms. }
    \label{fig:double_broad_double_peaked}
\end{figure*}

\begin{deluxetable*}{lc rl rl rl rl r}
\tabletypesize{\scriptsize}
\tablecaption{Double-peaked narrow emission-line BAT AGNs}
\tablewidth{0pt}
\tablecolumns{22}
\tablehead{
\colhead{ID\tablenotemark{a}} & 
\colhead{$\Delta$V\tablenotemark{b}}  &
\colhead{${\rm \lambda}_{\rm H\beta, b}$\tablenotemark{c}} &
\colhead{${\rm \lambda}_{\rm H\beta, r}$} &
\colhead{${\rm F}_{\rm H\beta, b}$\tablenotemark{d}} &
\colhead{${\rm F}_{\rm H\beta, r}$} &
\colhead{${\rm \lambda}_{\rm [OIII], b}$} &
\colhead{${\rm \lambda}_{\rm [OIII], r}$} &
\colhead{${\rm F}_{\rm [OIII], b}$} &
\colhead{${\rm F}_{\rm [OIII], r}$} &
\colhead{${\rm FWHM}_{\rm f}$\tablenotemark{e}} \\
\colhead{} &
\colhead{} &
\colhead{${\rm \lambda}_{\rm H\alpha, b}$} &
\colhead{${\rm \lambda}_{\rm H\alpha, r}$} &
\colhead{${\rm F}_{\rm H\alpha, b}$} &
\colhead{${\rm F}_{\rm H\alpha, r}$} &
\colhead{${\rm \lambda}_{\rm [NII], b}$} &
\colhead{${\rm \lambda}_{\rm [NII], r}$} &
\colhead{${\rm F}_{\rm [NII], b}$} &
\colhead{${\rm F}_{\rm [NII], r}$} &
\colhead{${\rm FWHM}_{\rm B}$\tablenotemark{e}} 
}
\startdata
  37&   173&       $-1.70$&$1.20$&            $51.5\pm5.0$&$49.7\pm4.8$&       $-1.70$&$1.20$&        $590.8\pm46.3$&$610.8\pm59.4$&      82\\
    &      &       $-2.10$&$1.70$&        $202.5\pm14.2$&$197.3\pm14.1$&       $-2.10$&$1.70$&          $116.1\pm8.4$&$107.1\pm8.0$&      82\\
  87&   168&        $0.00$&$3.40$&                      $<19.4$&$<19.4$&             $-0.25$&&                     $218.0\pm240.4$&&      94\\
    &      &        $1.50$&$5.50$&          $50.0\pm54.3$&$85.9\pm67.5$&        $1.70$&$5.00$&        $138.7\pm125.4$&$97.9\pm78.5$&      94\\
  89&   191&       $-2.00$&$2.00$&          $112.2\pm9.2$&$112.6\pm7.8$&       $-2.00$&$2.00$&        $715.1\pm48.8$&$875.8\pm59.7$&     157\\
    &      &       $-2.40$&$1.80$&        $358.9\pm26.2$&$410.2\pm25.7$&       $-2.40$&$1.80$&        $141.7\pm22.5$&$198.7\pm27.5$&     157\\
 159&   283&       $-3.00$&$3.00$&         $253.8\pm6.8$&$144.6\pm10.8$&       $-3.00$&$3.00$&       $1254.8\pm17.8$&$392.5\pm39.2$&     113\\
    &      &       $-3.40$&$2.80$&        $731.4\pm19.5$&$287.3\pm52.2$&       $-3.40$&$2.80$&        $322.3\pm13.9$&$191.0\pm23.6$&     113\\
 305&   274&       $-6.00$&$2.20$&        $170.7\pm11.8$&$245.7\pm35.7$&       $-6.00$&$2.20$&       $422.6\pm25.3$&$1094.1\pm77.0$&     164\\
    &      &       $-6.00$&$0.00$&      $107.3\pm13.4$&$2059.6\pm102.0$&       $-6.00$&$0.00$&        $116.1\pm13.3$&$993.9\pm49.2$&     164\\
 442&   329&      $-4.50$&$-0.60$&                 $88.8\pm6.8$&$<38.9$&      $-4.50$&$-0.60$&        $831.0\pm72.4$&$496.3\pm42.5$&     141\\
    &      &       $-7.00$&$0.20$&        $256.5\pm19.3$&$126.1\pm12.5$&       $-7.00$&$0.20$&         $63.0\pm15.3$&$208.2\pm16.8$&     141\\
 489&   114&       $-1.50$&$1.70$&          $104.3\pm6.9$&$68.5\pm11.5$&       $-1.50$&$1.70$&    $2930.6\pm172.2$&$3225.1\pm163.3$&      56\\
    &      &       $-1.00$&$1.50$&        $298.0\pm19.7$&$247.5\pm21.8$&       $-1.00$&$1.50$&        $294.3\pm17.9$&$548.8\pm28.9$&      56\\
 823&   185&       $-3.60$&$1.65$&        $1125.4\pm20.1$&$258.0\pm7.0$&             $-3.60$&&                     $6921.7\pm89.3$&&     141\\
    &      &      $-4.60$&$-0.55$&      $3586.3\pm57.5$&$1567.5\pm38.9$&      $-4.60$&$-0.55$&      $4978.8\pm60.6$&$1996.3\pm31.0$&     141\\
 970&   328&       $-2.70$&$2.70$&           $80.2\pm3.9$&$183.9\pm6.5$&       $-2.70$&$2.70$&      $1126.9\pm28.8$&$2322.2\pm53.7$&      42\\
    &      &       $-3.70$&$3.50$&        $254.7\pm11.2$&$959.8\pm20.9$&       $-3.70$&$3.50$&        $354.5\pm13.0$&$809.8\pm19.9$&      42\\
 986&   228&        $0.00$&$5.00$&        $978.9\pm34.0$&$857.7\pm12.4$&              $0.00$&&                     $3385.5\pm53.8$&&     246\\
    &      &        $0.00$&$5.00$&     $5827.2\pm97.3$&$7193.5\pm109.1$&        $0.00$&$5.00$&      $4289.9\pm59.5$&$3751.7\pm50.6$&     282\\
1072&   347&       $-4.90$&$1.30$&           $34.5\pm5.4$&$79.5\pm14.5$&       $-4.90$&$1.30$&       $424.6\pm18.6$&$1185.6\pm52.6$&     188\\
    &      &       $-5.90$&$1.70$&        $111.7\pm15.3$&$133.0\pm76.1$&             $-5.90$&&                       $253.2\pm8.9$&&     188\\
1139&   159&       $-2.50$&$1.00$&                $<48.6$&$117.9\pm7.6$&       $-2.50$&$1.00$&      $5995.5\pm21.8$&$3318.8\pm27.0$&     117\\
    &      &       $-2.50$&$1.00$&       $1170.5\pm25.2$&$765.7\pm18.3$&       $-2.50$&$1.00$&        $757.6\pm14.9$&$254.1\pm18.2$&     117\\
1150&   233&       $-2.80$&$2.30$&          $443.1\pm9.8$&$211.9\pm9.7$&             $-0.50$&&                      $469.8\pm13.0$&&      37\\
    &      &       $-2.80$&$2.30$&      $1437.2\pm28.0$&$1956.8\pm20.3$&       $-2.80$&$2.30$&        $750.4\pm13.4$&$650.7\pm17.3$&      37\\
1167&   333&       $-3.50$&$3.80$&                $62.1\pm12.1$&$<35.7$&       $-3.50$&$3.80$&        $296.2\pm52.0$&$210.8\pm38.9$&     211\\
    &      &       $-3.50$&$3.80$&        $222.2\pm34.5$&$124.2\pm21.1$&       $-3.50$&$3.80$&          $81.8\pm13.3$&$65.1\pm12.8$&     211\\
1174&   123&       $-1.20$&$3.00$&                 $29.3\pm5.3$&$<24.7$&       $-1.20$&$3.00$&          $214.5\pm9.8$&$208.1\pm6.2$&     176\\
    &      &       $-1.20$&$1.50$&        $102.7\pm15.2$&$335.6\pm15.8$&       $-1.20$&$1.50$&         $153.1\pm22.4$&$229.2\pm8.1$&     149\\
1180&   132&        $0.00$&$2.90$&        $933.5\pm17.3$&$316.0\pm17.6$&              $0.50$&&                     $4355.2\pm67.7$&&      30\\
    &      &        $0.00$&$2.90$&     $3125.1\pm49.4$&$1496.9\pm282.5$&        $0.00$&$2.90$&      $1735.3\pm24.0$&$1328.2\pm21.8$&      56\\
1186&   205&       $-2.50$&$1.50$&          $51.5\pm12.5$&$46.2\pm14.6$&       $-2.50$&$1.50$&       $169.6\pm42.6$&$744.0\pm188.7$&      45\\
    &      &       $-2.50$&$2.00$&        $152.4\pm35.7$&$279.2\pm54.4$&       $-2.50$&$2.00$&       $259.2\pm51.9$&$592.4\pm110.6$&      59\\
\enddata
\label{tab:double_peaked}
\tablenotetext{a}{Swift-BAT 70-month hard X-ray survey ID (\url{http://swift.gsfc.nasa.gov/results/bs70mon/}).}
\tablenotetext{b}{Average velocity offset in units of \kms\ between double-peaked narrow emission-lines measured from \NII\ and \Ha.}
\tablenotetext{c}{Offset in the two emission lines (blue and red component, in units of \AA) measured from \Hb\ ($4861.32$ \AA) in the rest-frame. For \OIII, \Ha, and \NII, $5006.77$ \AA, $6562.8$ \AA, and $6583.34$ \AA\ is used, respectively.}
\tablenotetext{d}{Emission line flux in units of $10^{-17}~{\rm erg}~{\rm cm}^{-2}~{\rm s}^{-1}$.}
\tablenotetext{e}{FWHM of forbidden lines (`f') and Balmer lines (`B') in units of \kms.}
\end{deluxetable*}

\subsection{Comparison with the optically selected AGNs from the SDSS}
We present comparisons between BAT AGNs and optically selected SDSS AGNs using the OSSY catalog \citep{Oh11, Oh15} in Figure~\ref{fig:LOIII_z_Balmer_decrement}. The BAT AGNs exhibit higher \OIII\ luminosities than those of the SDSS AGNs at any given redshift below 0.2, regardless of the AGN type. The \OIII\ luminosities of BAT AGNs are on average 0.79 dex (broad-line AGNs) and 0.73 dex (narrow-line AGNs) higher than that of the SDSS AGNs. Notably, the AGN-type classifications of the OSSY catalog used in Figure~\ref{fig:LOIII_z_Balmer_decrement} are based on the presence of broad \Ha\ emission lines. We also find that BAT narrow-line AGNs are dustier than SDSS narrow-line AGNs that exhibit high Balmer decrements (e.g., ${\rm H}\alpha/{\rm H}\beta_{\rm obs.} > 5.0$, $\sim36\%$). A high fraction of dusty AGNs selected using hard X-rays implies that optical selection is not ideal for the study of the most obscured and dusty AGNs.

\begin{deluxetable*}{lcccccccccccc}
\tabletypesize{\scriptsize}
\tablecaption{Wing components} 
\tablecolumns{13}
\tablehead{
\colhead{ID\tablenotemark{a}} & 
\colhead{$\lambda_{\rm H\beta}$\tablenotemark{b}} & 
\colhead{${\rm FWHM}_{\rm H\beta}$\tablenotemark{c}} & 
\colhead{${\rm F}_{\rm H\beta}$\tablenotemark{d}} &
\colhead{$\lambda_{\rm [OIII]}$} & 
\colhead{${\rm FWHM}_{\rm [OIII]}$} & 
\colhead{${\rm F}_{\rm [OIII]}$} &
\colhead{$\lambda_{\rm H\alpha}$} & 
\colhead{${\rm FWHM}_{\rm H\alpha}$} & 
\colhead{${\rm F}_{\rm H\alpha}$} &
\colhead{$\lambda_{\rm [NII]}$} & 
\colhead{${\rm FWHM}_{\rm [NII]}$} & 
\colhead{${\rm F}_{\rm [NII]}$}
}
\startdata
6 &  $\cdot \cdot \cdot$\tablenotemark{e}&  $\cdot \cdot \cdot$&  $\cdot \cdot \cdot$&               5003.6&                  447&      $3383.2\pm11.4$&  $\cdot \cdot \cdot$&  $\cdot \cdot \cdot$&  $\cdot \cdot \cdot$&  $\cdot \cdot \cdot$&  $\cdot \cdot \cdot$&  $\cdot \cdot \cdot$\\
33 &  $\cdot \cdot \cdot$&  $\cdot \cdot \cdot$&  $\cdot \cdot \cdot$&               5003.7&                  436&        $805.7\pm0.2$&  $\cdot \cdot \cdot$&  $\cdot \cdot \cdot$&  $\cdot \cdot \cdot$&  $\cdot \cdot \cdot$&  $\cdot \cdot \cdot$&  $\cdot \cdot \cdot$\\
43 &  $\cdot \cdot \cdot$&  $\cdot \cdot \cdot$&  $\cdot \cdot \cdot$&               5003.3&                  493&        $326.8\pm2.1$&  $\cdot \cdot \cdot$&  $\cdot \cdot \cdot$&  $\cdot \cdot \cdot$&  $\cdot \cdot \cdot$&  $\cdot \cdot \cdot$&  $\cdot \cdot \cdot$\\
55 &  $\cdot \cdot \cdot$&  $\cdot \cdot \cdot$&  $\cdot \cdot \cdot$&               5012.7&                  829&       $2440.9\pm7.0$&  $\cdot \cdot \cdot$&  $\cdot \cdot \cdot$&  $\cdot \cdot \cdot$&  $\cdot \cdot \cdot$&  $\cdot \cdot \cdot$&  $\cdot \cdot \cdot$\\
60 &  $\cdot \cdot \cdot$&  $\cdot \cdot \cdot$&  $\cdot \cdot \cdot$&               4998.4&                 1177&       $3626.9\pm0.3$&  $\cdot \cdot \cdot$&  $\cdot \cdot \cdot$&  $\cdot \cdot \cdot$&  $\cdot \cdot \cdot$&  $\cdot \cdot \cdot$&  $\cdot \cdot \cdot$\\
61 &  $\cdot \cdot \cdot$&  $\cdot \cdot \cdot$&  $\cdot \cdot \cdot$&               5004.2&                  365&        $975.3\pm4.6$&  $\cdot \cdot \cdot$&  $\cdot \cdot \cdot$&  $\cdot \cdot \cdot$&  $\cdot \cdot \cdot$&  $\cdot \cdot \cdot$&  $\cdot \cdot \cdot$\\
76 &  $\cdot \cdot \cdot$&  $\cdot \cdot \cdot$&  $\cdot \cdot \cdot$&               5001.9&                  680&       $330.5\pm10.0$&  $\cdot \cdot \cdot$&  $\cdot \cdot \cdot$&  $\cdot \cdot \cdot$&  $\cdot \cdot \cdot$&  $\cdot \cdot \cdot$&  $\cdot \cdot \cdot$\\
78 &  $\cdot \cdot \cdot$&  $\cdot \cdot \cdot$&  $\cdot \cdot \cdot$&               5004.2&                  367&       $2054.0\pm0.1$&  $\cdot \cdot \cdot$&  $\cdot \cdot \cdot$&  $\cdot \cdot \cdot$&  $\cdot \cdot \cdot$&  $\cdot \cdot \cdot$&  $\cdot \cdot \cdot$\\
79 &  $\cdot \cdot \cdot$&  $\cdot \cdot \cdot$&  $\cdot \cdot \cdot$&               5003.6&                  448&       $1254.8\pm3.9$&  $\cdot \cdot \cdot$&  $\cdot \cdot \cdot$&  $\cdot \cdot \cdot$&  $\cdot \cdot \cdot$&  $\cdot \cdot \cdot$&  $\cdot \cdot \cdot$\\
80 &  $\cdot \cdot \cdot$&  $\cdot \cdot \cdot$&  $\cdot \cdot \cdot$&               4999.6&                 1017&       $309.2\pm35.1$&  $\cdot \cdot \cdot$&  $\cdot \cdot \cdot$&  $\cdot \cdot \cdot$&  $\cdot \cdot \cdot$&  $\cdot \cdot \cdot$&  $\cdot \cdot \cdot$\\
81 &  $\cdot \cdot \cdot$&  $\cdot \cdot \cdot$&  $\cdot \cdot \cdot$&               5001.8&                  705&       $163.0\pm31.0$&  $\cdot \cdot \cdot$&  $\cdot \cdot \cdot$&  $\cdot \cdot \cdot$&  $\cdot \cdot \cdot$&  $\cdot \cdot \cdot$&  $\cdot \cdot \cdot$\\
83 &  $\cdot \cdot \cdot$&  $\cdot \cdot \cdot$&  $\cdot \cdot \cdot$&               5001.8&                  706&      $1089.0\pm26.9$&  $\cdot \cdot \cdot$&  $\cdot \cdot \cdot$&  $\cdot \cdot \cdot$&  $\cdot \cdot \cdot$&  $\cdot \cdot \cdot$&  $\cdot \cdot \cdot$\\
87 &               4861.7&                  209&              $<43.1$&               5009.4&                  164&      $214.5\pm226.4$&               6561.1&                  117&      $210.9\pm166.4$&               6581.6&                  117&      $248.0\pm243.2$\\
89 &               4864.9&                  807&       $148.7\pm21.5$&               5011.4&                  942&     $1552.1\pm100.1$&               6565.6&                  537&       $511.1\pm46.1$&               6584.6&                  370&       $197.3\pm55.7$\\
98 &  $\cdot \cdot \cdot$&  $\cdot \cdot \cdot$&  $\cdot \cdot \cdot$&               5011.3&                  633&        $226.1\pm0.8$&  $\cdot \cdot \cdot$&  $\cdot \cdot \cdot$&  $\cdot \cdot \cdot$&  $\cdot \cdot \cdot$&  $\cdot \cdot \cdot$&  $\cdot \cdot \cdot$\\
\enddata
\tablenotetext{a}{Swift-BAT 70-month hard X-ray survey ID (\url{http://swift.gsfc.nasa.gov/results/bs70mon/}).}
\tablenotetext{b}{Emission line center in units of \AA.}
\tablenotetext{c}{FWHM in units of \kms.}
\tablenotetext{d}{Emission line flux in units of $10^{-17}~{\rm erg}~{\rm cm}^{-2}~{\rm s}^{-1}$.}
\tablecomments{(This table is available in its entirety in a machine-readable form in the online journal. A portion is shown here for guidance regarding its form and content.)}
\label{tab:wings}
\end{deluxetable*}

\subsection{Complex emission-line features}

The emission line profiles of AGNs are quite complex exhibiting asymmetric broad-line features (\citealt{Sulentic89, Marziani96, Sulentic00}, and references therein). The line profiles are superposition of different components such as Doppler motions, outflowing gas, turbulent motions in the extended accretion disks, and electron scattering \citep{Laor06, Kollatschny13}. These various components result in different emission line shapes can be described as Gaussian, Lorentzian, exponential, and logarithmic profiles. 

In order to explain the complex observational features of the broad Balmer regions, we have fitted the optical spectra of BAT AGNs using multiple Gaussian components \citep{Mullaney08, LaMura09, Suh15, Oh19, Suh20}. Several previous studies have suggested a possible physical origin of complex broad-lines (and/or double-peaked narrow emission lines), including rotating disks, a binary broad-line region in a binary supermassive black hole system, complex narrow-line region kinematics, and biconical outflows \citep{Gaskell83, Chen89, Zheng90, Eracleous94, Eracleous09, Shen11}; however, this remains an open question. We report examples of BAT AGNs that display complex broad-line features ($N=250$) in Figure~\ref{fig:double_broad_double_peaked} and provide properties of double-peaked narrow emission lines ($N=17$) in Table~\ref{tab:double_peaked}. 

Previous studies demonstrated that $1\%$ of type 2 AGNs at $z\sim0.1$ present double-peaked narrow emission lines with a velocity splitting of a few hundred \kms\ \citep{Wang09, Liu10}, which is good agreement with our result ($2\%$, $17/743$). However, our high-resolution high S/N (X-Shooter) data, which comprises $23\%$ ($\NXshooter/\Ntotalobtained$) of the sample, had a $10\%$ ($17/\NXshooter$) double-peaked narrow emission line fraction. Thus ours ($2\%$, $17/743$)  should be regarded as a lower limit in the context of the spectral inhomogeneity of the survey. This is consistent with a recent study by \citet{Lyu16}, which found that double-peaked narrow-emission-line AGN made up $\sim9\%$ of AGN selected from the SDSS DR10.

Table~\ref{tab:wings} presents emission line centers, FWHMs, and fluxes of the wing components applied to \Hb, \OIII, \Ha, and \NIIb\ emission lines.

\begin{figure} 
	\centering
	\includegraphics[width=0.99\linewidth]{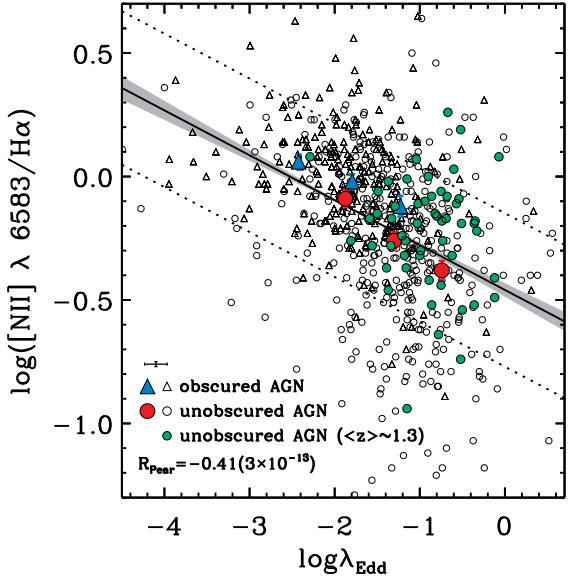} 
    \caption{Optical emission-line ratio (\NIIHa) vs. Eddington ratio ($\lambda_{\rm Edd}$) diagram. Black open circles and triangles indicate unobscured AGNs (i.e., Sy1, Sy1.2, Sy1.5, Sy1.8, and Sy1.9) and obscured AGNs (Sy2), respectively. The median in each bin of these AGNs is indicated by blue filled triangles and red filled circles. The black solid line indicates the measured anti-correlation. The grey-shaded regions account for the errors in the slope and intercept. The dotted lines indicate RMS deviation. Green filled circles are unobscured AGNs at a higher redshift at $ \langle z \rangle = 1.3$ \citep{Oh19}. Pearson correlation coefficient, $p$-value, and typical uncertainties are shown in the bottom-left corner.}
    \label{fig:NIIHa_Edd_ratio}
\end{figure}

\subsection{\NIIHa\ versus Eddington ratio}
The observed relationship between the AGN Eddington ratio ($\lambda_{\rm Edd}$) and the optical narrow-emission-line ratio, \NIIHa, was investigated using X-ray-selected AGNs. \citet{Oh17} showed an anti-correlation using 297 nearby BAT AGNs, which is explained by X-ray-heating processes and/or the presence of radiatively driven outflows in the high-$\lambda_{\rm Edd}$ state. The observed anti-correlation still holds in the higher redshift regime, up to $z\sim1.7$ \citep[][green filled dots in Figure~\ref{fig:NIIHa_Edd_ratio}]{Oh19}. We show the anti-correlation in Figure~\ref{fig:NIIHa_Edd_ratio} using the measurements of black hole mass \citep{Koss_DR2_sigs,Mejia_Broadlines}, bolometric luminosity \citep{Ricci17}, and the \NIIHa\ ratio of 639 BAT AGNs. We report $\alpha$ and $\beta$ values of $-0.46\pm0.03$ and $-0.18\pm0.02$, respectively, where $\alpha$ and $\beta$ are the intercept and slope of the Bayesian linear regression as follows: 
\begin{equation}
	  \log(F_{\rm line}/F_{\rm Balmer})= \alpha + \beta\log\lambda_{\rm Edd} 	
\label{eq:regression}
\end{equation} 
The root mean square deviation is 0.31 dex, which is comparable to that obtained by \citealt{Oh17} (0.28 dex, cf. $\alpha=-0.42\pm0.04$, $\beta=-0.19\pm0.02$). The Pearson $R$-coefficient and $p$-value are $-0.41\pm0.05$ and $3\times10^{-13}$, respectively, which reassures the statistical significance, as shown in a study by \citet{Oh17}.

\section{Summary and Conclusions}
\label{sec:summary}

We presented the second data release of the BAT AGN spectroscopic survey, ultra-hard X-ray-selected nearby, powerful AGNs, and the optical spectroscopic follow-up project conducted with dedicated observation campaigns and public archival data. The key features of this study compared with the first data release are as follows:   

\begin{enumerate}
\item The DR2 emission line datasets comprise $\Ntotalobtained$ high quality optical spectra, which is $99.6\%$ of the non-beamed and unlensed AGNs from the Swift-BAT 70-month ultra-hard X-ray all-sky survey catalog.
\item Spectral incompleteness, such as insufficient spectral coverage and/or low S/N, decreased below $2\%$ ($14/\Ntotalobtained$), which enabled an investigation of the optical spectroscopic properties of unexplored BAT AGNs. 
\end{enumerate}

Our main findings are as follows: 
\begin{enumerate}
	 \item AGN subtypes that were classified using optical emission-line analysis are in good agreement with X-ray obscuration, and they exhibit a dichotomy at $N_{\rm H}=10^{22}~{\rm cm}~{\rm s}^{-2}$. 
	 \item The type 1 AGN fraction, both with broad \Hb\ and/or \Ha, increases with increasing $2$-$10$ keV intrinsic luminosity. 
	 \item The most commonly used emission-line diagnostic diagram, \OIII/\Hb\ versus \NII/\Ha, yields a $75.4\%$ ($560/\Ntotalobtained$) fraction of the Seyfert class; however, only a few percent were assigned to the LINERs ($4.6\%$), composite ($5.5\%$), and HII ($2.3\%$) classes. Owing to difficulties in the line detection of \OIIa\ and \HeII, \OIII/\OIIa\ versus \NII/\Ha\ and \HeII/\Hb\ versus \NII/\Ha\ diagrams exhibit lower detection rates with higher fractions of the `non-detection' class. However, the overall trend was consistent with the dominant fraction of the Seyfert class.
	  \item Compared with optically selected narrow-line AGNs in the SDSS, the X-ray-selected BAT AGNs shown in this study present a higher fraction of dustier galaxies with \Ha/\Hb$>5$. Moreover, BAT AGNs exhibit higher \OIII\ luminosity than SDSS AGNs, regardless of the presence of broad Balmer lines across the considered redshift range.
	  \item We present a subpopulation of AGNs that feature complex broad-line emissions ($34\%$, $250/743$) or double-peaked narrow lines ($2\%$, $17/743$). 
	  \item  An anti-correlation between the AGN Eddington ratio and optical narrow-emission-line ratio is observed for more than double the number of BAT AGNs compared with the previous study.
\end{enumerate}

We provide all optical spectra and best fits with measured quantities to the community through the BASS Website so that the database may be useful for many fruitful science applications. 

\begin{acknowledgments}
K.O. acknowledges support from the National Research Foundation of Korea (NRF-2020R1C1C1005462).
This research was supported in part by the Japan Society for the Promotion of Science (JSPS) postdoctoral fellowship program for research in Japan (ID: P17321). 
We acknowledge support from NASA through ADAP award NNH16CT03C (M.K.); the JSPS KAKENHI Grant Number 20H01946 (Y.U.); the Fondecyt Iniciacion grant 11190831 (C.R.); the Israel Science Foundation (grant number 1849/19, B.T.); the Comunidad de Madrid through the Atracci\'on de Talento Investigador Grant 2018-T1/TIC-11035 (I.L.); FONDECYT Postdoctorado project No. 3210157 (A.R.L.). R.R. thanks to Conselho Nacional de Desenvolvimento Cient\'{i}fico e
Tecnol\'ogico  (CNPq, Proj. 311223/2020-6,  304927/2017-1 and
400352/2016-8), Funda\c{c}\~ao de amparo 'a pesquisa do Rio Grande do
Sul (FAPERGS, Proj. 16/2551-0000251-7 and 19/1750-2),
Coordena\c{c}\~ao de Aperfei\c{c}oamento de Pessoal de N\'{i}vel Superior (CAPES, Proj. 0001). 
We acknowledge support from FONDECYT Regular 1190818 (E.T., F.E.B.) and 1200495 (F.E.B., E.T.), ANID grants CATA-Basal AFB-170002 and FB210003 (E.T., F.E.B.), Anillo ACT172033 (E.T.), Millennium Nucleus NCN19\_058 (TITANs; E.T.)
and Millennium Science Initiative Program  - ICN12\_009 (MAS; F.E.B.). 
C.M.U. acknowledges support from the National Science Foundation under Grant No. AST-1715512.

Some of the optical spectra were taken with Doublespec at Palomar via Yale (PI M. Powell, 2017-2019, 16 nights) as well as Caltech (PI F. Harrison) and JPL (PI D. Stern) from programs from 2013-2020.

This work made use of observations collected at the European Southern Observatory under ESO programs 98.A-0635, 99-A-0403, 100.B-0672, 101.A-0765, 102.A-0433, 103.A-0521, and 104.A-0353.  Based on observations from seven CNTAC programs:  CN2016A-80, CN2018A-104, CN2018B-83, CN2019A-70, CN2019B-77, CN2020A-90,  and CN2020B-48 (PI C. Ricci).  Based on observations obtained at the Southern Astrophysical Research (SOAR) telescope, which is a joint project of the Minist\'{e}rio da Ci\^{e}ncia, Tecnologia e Inova\c{c}\~{o}es (MCTI/LNA) do Brasil, the US National Science Foundation’s NOIRLab, the University of North Carolina at Chapel Hill (UNC), and Michigan State University (MSU).  Based on observations at Kitt Peak National Observatory at NSF’s NOIRLab (NOIRLab Prop. ID 52, 2946; PI: F. Bauer), which is managed by the Association of Universities for Research in Astronomy (AURA) under a cooperative agreement with the National Science Foundation.  Some of the data presented herein were obtained at the W.M. Keck Observatory, which is operated as a scientific partnership among the California Institute of Technology, the University of California and the National Aeronautics and Space Administration. The Observatory was made possible by the generous financial support of the W.M. Keck Foundation.  The authors wish to recognize and acknowledge the very significant cultural role and reverence that the summit of Mauna Kea has always had within the indigenous Hawaiian community.  We are most fortunate to have the opportunity to conduct observations from this mountain. 

This research has made use of NASA's ADS Service. 
This research has made use of the NASA/ IPAC Infrared Science Archive, which is operated by the Jet Propulsion Laboratory, California Institute of Technology, under contract with the National Aeronautics and Space Administration.

Funding for SDSS-III has been provided by the Alfred P. Sloan Foundation, the Participating Institutions, the National Science Foundation, and the U.S. Department of Energy Office of Science. The SDSS-III web site is http://www.sdss3.org/. SDSS-III is managed by the Astrophysical Research Consortium for the Participating Institutions of the SDSS-III Collaboration including the University of Arizona, the Brazilian Participation Group, Brookhaven National Laboratory, Carnegie Mellon University, University of Florida, the French Participation Group, the German Participation Group, Harvard University, the Instituto de Astrofisica de Canarias, the Michigan State/Notre Dame/JINA Participation Group, Johns Hopkins University, Lawrence Berkeley National Laboratory, Max Planck Institute for Astrophysics, Max Planck Institute for Extraterrestrial Physics, New Mexico State University, New York University, Ohio State University, Pennsylvania State University, University of Portsmouth, Princeton University, the Spanish Participation Group, University of Tokyo, University of Utah, Vanderbilt University, University of Virginia, University of Washington, and Yale University.
 
The Digitized Sky Surveys were produced at the Space Telescope Science Institute under U.S. Government grant NAG W-2166. The images of these surveys are based on photographic data obtained using the Oschin Schmidt Telescope on Palomar Mountain and the UK Schmidt Telescope. The plates were processed into the present compressed digital form with the permission of these institutions. The National Geographic Society - Palomar Observatory Sky Atlas (POSS-I) was made by the California Institute of Technology with grants from the National Geographic Society. The Second Palomar Observatory Sky Survey (POSS-II) was made by the California Institute of Technology with funds from the National Science Foundation, the National Geographic Society, the Sloan Foundation, the Samuel Oschin Foundation, and the Eastman Kodak Corporation. The Oschin Schmidt Telescope is operated by the California Institute of Technology and Palomar Observatory. The UK Schmidt Telescope was operated by the Royal Observatory Edinburgh, with funding from the UK Science and Engineering Research Council (later the UK Particle Physics and Astronomy Research Council), until 1988 June, and thereafter by the Anglo-Australian Observatory. The blue plates of the southern Sky Atlas and its Equatorial Extension (together known as the SERC-J), as well as the Equatorial Red (ER), and the Second Epoch [red] Survey (SES) were all taken with the UK Schmidt.
\end{acknowledgments}

\begin{rotatetable*}
\begin{deluxetable*}{lcc ccccc ccc cc}
\tabletypesize{\scriptsize}
\tablecaption{Emission-line Measurements $\textendash$ from HeII~$\lambda$3203 to H$\gamma$~$\lambda$4340} 
\tablecolumns{13}
\tablehead{
\colhead{ID\tablenotemark{a}} & 
\colhead{HeII\tablenotemark{b}} & 
\colhead{[NeV]\tablenotemark{b}}  &
\colhead{[NeV]\tablenotemark{b}} &
\colhead{[OII]\tablenotemark{b}} &
\colhead{[NeIII]\tablenotemark{b}} &
\colhead{[NeIII]\tablenotemark{b}} &
\colhead{H$\zeta$\tablenotemark{b}} &
\colhead{H$\epsilon$\tablenotemark{b}} &
\colhead{H$\delta$\tablenotemark{b}} &
\colhead{H$\gamma$\tablenotemark{b}} &
\colhead{FWHM\tablenotemark{c}} &
\colhead{flag\tablenotemark{d}}
\\
\colhead{} &
\colhead{$\lambda$3203} &
\colhead{$\lambda$3345} &
\colhead{$\lambda$3425} &
\colhead{$\lambda$3727} &
\colhead{$\lambda$3868} &
\colhead{$\lambda$3967} &
\colhead{$\lambda$3889} &
\colhead{$\lambda$3970} &
\colhead{$\lambda$4101} &
\colhead{$\lambda$4340} &
\colhead{} &
\colhead{}
}
\startdata
  1 &  $\cdot \cdot \cdot$\tablenotemark{e} &  $\cdot \cdot \cdot$ &  $\cdot \cdot \cdot$ &        $185.8\pm1.8$ &        $101.6\pm1.0$ &         $20.1\pm0.3$ &          $9.6\pm0.2$ &         $15.1\pm0.2$ &         $26.0\pm0.4$ &         $51.6\pm0.7$ &            $212\pm9$ & n\\
2 &  $\cdot \cdot \cdot$ &  $\cdot \cdot \cdot$ &  $\cdot \cdot \cdot$ &       $537.8\pm25.0$ &       $309.3\pm13.3$ &        $157.4\pm7.4$ &              $<88.6$ &              $<90.5$ &         $83.2\pm4.7$ &        $162.8\pm8.4$ &           $657\pm15$ & y\\
3 &  $\cdot \cdot \cdot$ &  $\cdot \cdot \cdot$ &  $\cdot \cdot \cdot$ &  $\cdot \cdot \cdot$ &  $\cdot \cdot \cdot$ &  $\cdot \cdot \cdot$ &  $\cdot \cdot \cdot$ &  $\cdot \cdot \cdot$ &  $\cdot \cdot \cdot$ &  $\cdot \cdot \cdot$ &           $586\pm27$ & y\\
4 &  $\cdot \cdot \cdot$ &  $\cdot \cdot \cdot$ &  $\cdot \cdot \cdot$ &        $541.8\pm4.7$ &        $126.9\pm1.0$ &         $30.4\pm0.3$ &              $<25.2$ &              $<24.5$ &         $30.0\pm0.5$ &         $63.5\pm0.9$ &           $390\pm12$ & n\\
5 &  $\cdot \cdot \cdot$ &  $\cdot \cdot \cdot$ &       $5796.2\pm5.4$ &            $<4619.3$ &            $<4238.0$ &            $<4346.2$ &            $<4159.1$ &            $<4349.1$ &            $<4386.6$ &            $<1652.4$ &           $455\pm22$ & n\\
6 &  $\cdot \cdot \cdot$ &  $\cdot \cdot \cdot$ &  $\cdot \cdot \cdot$ &  $\cdot \cdot \cdot$ &  $\cdot \cdot \cdot$ &  $\cdot \cdot \cdot$ &  $\cdot \cdot \cdot$ &  $\cdot \cdot \cdot$ &  $\cdot \cdot \cdot$ &      $1626.0\pm11.5$ &           $447\pm15$ & y\\
7 &  $\cdot \cdot \cdot$ &  $\cdot \cdot \cdot$ &  $\cdot \cdot \cdot$ &       $1331.7\pm9.5$ &        $304.4\pm2.3$ &         $99.3\pm1.1$ &              $<79.7$ &              $<83.6$ &         $82.0\pm0.9$ &        $169.1\pm1.7$ &           $617\pm14$ & n\\
10 &              $<69.6$ &              $<72.7$ &        $58.9\pm16.9$ &        $83.7\pm17.7$ &              $<45.6$ &              $<46.8$ &              $<41.8$ &              $<46.8$ &              $<44.1$ &              $<37.9$ &           $275\pm20$ & n\\
13 &  $\cdot \cdot \cdot$ &  $\cdot \cdot \cdot$ &  $\cdot \cdot \cdot$ &  $\cdot \cdot \cdot$ &  $\cdot \cdot \cdot$ &  $\cdot \cdot \cdot$ &  $\cdot \cdot \cdot$ &  $\cdot \cdot \cdot$ &  $\cdot \cdot \cdot$ &  $\cdot \cdot \cdot$ &            $313\pm7$ & n\\
14 &         $97.2\pm7.9$ &              $<71.5$ &       $135.2\pm11.2$ &       $643.6\pm49.9$ &       $320.8\pm23.2$ &         $95.9\pm7.6$ &         $18.4\pm2.5$ &         $29.1\pm3.8$ &         $50.5\pm6.2$ &       $102.1\pm11.1$ &           $657\pm29$ & y\\
\enddata
\tablenotetext{a}{Swift-BAT 70-month hard X-ray survey ID (\url{http://swift.gsfc.nasa.gov/results/bs70mon/}).}
\tablenotetext{b}{Emission line flux in units of $10^{-17}~{\rm erg}~{\rm cm}^{-2}~{\rm s}^{-1}$.}
\tablenotetext{c}{FWHM for forbidden lines measured from \NIIb\ in units of \kms.}
\tablenotetext{d}{Flag indicating the use of broad Balmer components (${\rm H\delta, H\gamma, H\beta,~and~H\alpha}$) in spectral line fitting.}
\tablenotetext{e}{Symbols `$\cdot \cdot \cdot$' and `$<$' indicate a lack of spectral coverage and the $3\sigma$ upper limit estimation, respectively.}
\tablecomments{(This table is available in its entirety in a machine-readable form in the online journal. A portion is shown here for guidance regarding its form and content.)}
\label{tab:OIIcomplex}
\end{deluxetable*}
\end{rotatetable*}

\movetabledown=60mm
\begin{rotatetable*}
\begin{deluxetable*}{lcc ccccc ccc cc cc}
\tabletypesize{\scriptsize}
\tablecaption{Emission-line Measurements $\textendash$ from [OIII]~$\lambda$4363 to HeI~$\lambda$5875} 
\tablecolumns{15}
\tablehead{
\colhead{ID\tablenotemark{a}} & 
\colhead{[OIII]\tablenotemark{b}} & 
\colhead{HeII\tablenotemark{b}}  &
\colhead{[ArIV]\tablenotemark{b}} &
\colhead{[ArIV]\tablenotemark{b}} &
\colhead{H$\beta$\tablenotemark{b}} &
\colhead{[OIII]\tablenotemark{b}} &
\colhead{[OIII]\tablenotemark{b}} &
\colhead{[NI]\tablenotemark{b}} &
\colhead{[NI]\tablenotemark{b}} &
\colhead{HeI\tablenotemark{b}} &
\colhead{FWHM\tablenotemark{c}} &
\colhead{flag\tablenotemark{d}} &
\colhead{C\tablenotemark{e}} &
\colhead{$\frac{\rm H\alpha}{\rm H\beta}$}
\\
\colhead{} &
\colhead{$\lambda$4363} &
\colhead{$\lambda$4685} &
\colhead{$\lambda$4711} &
\colhead{$\lambda$4740} &
\colhead{$\lambda$4861} &
\colhead{$\lambda$4958} &
\colhead{$\lambda$5007} &
\colhead{$\lambda$5197} &
\colhead{$\lambda$5200} &
\colhead{$\lambda$5875} &
\colhead{} &
\colhead{} &
\colhead{} &
\colhead{} 
}
\startdata
1 &         $28.6\pm0.4$ &         $32.9\pm0.4$ &              $<15.7$\tablenotemark{f} &              $<15.7$ &        $133.1\pm1.4$ &        $459.0\pm4.3$ &      $1332.0\pm12.4$ &              $<17.3$ &              $<17.3$ &         $27.7\pm0.4$ &            $198\pm1$ &  n&                $1.39$ &                $4.51$\\
2 &        $154.4\pm6.5$ &        $148.3\pm5.7$ &         $52.8\pm3.2$ &        $111.5\pm4.6$ &       $407.9\pm18.0$ &      $1784.5\pm54.4$ &     $5166.4\pm155.4$ &         $57.8\pm8.9$ &         $66.9\pm9.8$ &        $161.5\pm5.9$ &           $657\pm27$ &  y&                $1.31$ &                $4.21$\\
3 &  $\cdot \cdot \cdot$ &  $\cdot \cdot \cdot$ &  $\cdot \cdot \cdot$ &  $\cdot \cdot \cdot$ &        $614.0\pm7.7$ &       $1031.5\pm3.3$ &       $2954.2\pm9.5$ &  $\cdot \cdot \cdot$ &      $2675.8\pm10.7$ &  $\cdot \cdot \cdot$ &           $586\pm14$ &  y&                $1.00$ &                $3.06$\\
4 &              $<26.9$ &         $37.2\pm0.4$ &              $<36.5$ &              $<36.8$ &        $185.3\pm2.0$ &        $530.9\pm3.2$ &       $1556.6\pm9.1$ &              $<40.3$ &              $<40.3$ &         $21.8\pm0.3$ &            $411\pm1$ &  n&                $1.80$ &                $6.09$\\
5 &            $<1701.5$ &       $1218.2\pm0.9$ &        $336.8\pm0.4$ &             $<386.8$ &       $1460.1\pm0.9$ &       $5195.7\pm3.3$ &      $14915.5\pm9.5$ &             $<424.2$ &  $\cdot \cdot \cdot$ &              $<82.6$ &           $445\pm13$ &  n&                $1.05$ &                $3.28$\\
6 &       $2049.3\pm8.4$ &      $3266.6\pm11.8$ &        $199.3\pm2.0$ &  $\cdot \cdot \cdot$ &      $4601.7\pm24.5$ &      $5104.2\pm17.3$ &     $14927.3\pm49.4$ &  $\cdot \cdot \cdot$ &  $\cdot \cdot \cdot$ &  $\cdot \cdot \cdot$ &           $447\pm25$ &  y&                $1.69$ &                $5.65$\\
7 &         $45.9\pm0.7$ &         $61.0\pm0.8$ &  $\cdot \cdot \cdot$ &  $\cdot \cdot \cdot$ &        $469.9\pm3.5$ &       $1370.9\pm7.9$ &      $4003.0\pm22.7$ &        $216.5\pm8.1$ &  $\cdot \cdot \cdot$ &        $115.2\pm4.7$ &            $601\pm2$ &  n&                $1.62$ &                $5.40$\\
10 &              $<41.8$ &              $<44.9$ &              $<45.2$ &  $\cdot \cdot \cdot$ &        $43.7\pm12.0$ &       $204.5\pm51.1$ &      $596.5\pm146.0$ &  $\cdot \cdot \cdot$ &  $\cdot \cdot \cdot$ &  $\cdot \cdot \cdot$ &           $250\pm10$ &  n&                $1.58$ &                $5.25$\\
13 &  $\cdot \cdot \cdot$ &            $<1048.1$ &            $<1053.8$ &  $\cdot \cdot \cdot$ &             $<990.4$ &       $3182.6\pm0.7$ &       $9664.0\pm2.1$ &  $\cdot \cdot \cdot$ &            $<1163.2$ &        $977.3\pm0.2$ &           $285\pm18$ &  n&   $\cdot \cdot \cdot$ &   $\cdot \cdot \cdot$\\
14 &         $98.7\pm6.9$ &        $114.8\pm8.2$ &              $<55.9$ &              $<56.3$ &       $272.7\pm23.7$ &      $1078.6\pm62.1$ &     $3139.2\pm177.4$ &         $53.7\pm9.4$ &              $<61.7$ &        $169.0\pm9.5$ &           $657\pm15$ &  y&                $1.49$ &                $4.91$\\
\enddata
\tablenotetext{a}{Swift-BAT 70-month hard X-ray survey ID (\url{http://swift.gsfc.nasa.gov/results/bs70mon/}).}
\tablenotetext{b}{Emission line flux in units of $10^{-17}~{\rm erg}~{\rm cm}^{-2}~{\rm s}^{-1}$.}
\tablenotetext{c}{FWHM for Balmer lines measured from \Ha\ in units of \kms.}
\tablenotetext{d}{Flag indicating the use of broad Balmer components (${\rm H\delta, H\gamma, H\beta,~and~H\alpha}$) in spectral line fitting.}
\tablenotetext{e}{Correction factor, where ${\rm [OIII]}_{\rm intr.}~=~C\times{\rm [OIII]}_{\rm obs.}$. We use the \citet{Cardelli89} reddening curve assuming an intrinsic ratio of $R_{\rm V}=3.1$ to correct for dust extinction.}
\tablenotetext{f}{Symbols `$\cdot \cdot \cdot$' and `$<$' indicate a lack of spectral coverage and the $3\sigma$ upper limit estimation, respectively.}
\tablecomments{(This table is available in its entirety in a machine-readable form in the online journal. A portion is shown here for guidance regarding its form and content.)}
\label{tab:OIIIcomplex}
\end{deluxetable*}
\end{rotatetable*}

\begin{rotatetable*}
\begin{deluxetable*}{lcc ccccc ccc c}
\tabletypesize{\scriptsize}
\tablecaption{Emission-line Measurements $\textendash$ from [OI]~$\lambda$6300 to [OII]~$\lambda$7330} 
\tablecolumns{12}
\tablehead{
\colhead{ID\tablenotemark{a}} & 
\colhead{[OI]\tablenotemark{b}} & 
\colhead{[OI]\tablenotemark{b}}  &
\colhead{[NII]\tablenotemark{b}} &
\colhead{H$\alpha$\tablenotemark{b}} &
\colhead{[NII]\tablenotemark{b}} &
\colhead{[SII]\tablenotemark{b}} &
\colhead{[SII]\tablenotemark{b}} &
\colhead{[ArIII]\tablenotemark{b}} &
\colhead{[OII]\tablenotemark{b}} &
\colhead{[OII]\tablenotemark{b}} &
\colhead{Flag\tablenotemark{c}} \\
\colhead{} &
\colhead{$\lambda$6300} &
\colhead{$\lambda$6363} &
\colhead{$\lambda$6548} &
\colhead{$\lambda$6562} &
\colhead{$\lambda$6583} &
\colhead{$\lambda$6716} &
\colhead{$\lambda$6730} &
\colhead{$\lambda$7135} &
\colhead{$\lambda$7319} &
\colhead{$\lambda$7330} &
\colhead{} 
}
\startdata
    1 &         $63.2\pm0.5$ &         $21.4\pm0.2$ &         $49.3\pm0.4$ &        $600.0\pm4.1$ &        $146.2\pm1.1$ &        $111.6\pm0.7$ &        $102.5\pm0.7$ &         $24.0\pm0.4$ &  $\cdot \cdot \cdot$\tablenotemark{d} &  $\cdot \cdot \cdot$ & y\\
2 &       $406.1\pm11.0$ &        $137.1\pm3.7$ &       $596.9\pm17.0$ &      $1716.9\pm51.5$ &      $1767.9\pm50.0$ &       $632.8\pm17.6$ &       $673.1\pm17.7$ &  $\cdot \cdot \cdot$ &  $\cdot \cdot \cdot$ &  $\cdot \cdot \cdot$ & y\\
3 &        $244.0\pm8.3$ &         $81.5\pm2.8$ &        $376.1\pm7.4$ &      $1879.1\pm21.9$ &      $1107.4\pm21.8$ &        $357.5\pm9.1$ &        $300.3\pm8.7$ &        $256.7\pm8.9$ &  $\cdot \cdot \cdot$ &  $\cdot \cdot \cdot$ & y\\
4 &        $114.8\pm0.6$ &         $38.6\pm0.2$ &        $195.9\pm1.0$ &       $1128.0\pm5.8$ &        $584.1\pm2.8$ &        $335.2\pm1.4$ &        $262.5\pm1.3$ &         $61.1\pm0.5$ &  $\cdot \cdot \cdot$ &  $\cdot \cdot \cdot$ & n\\
5 &        $452.8\pm0.4$ &        $151.5\pm0.1$ &       $1058.7\pm0.6$ &       $4793.2\pm2.6$ &       $3121.6\pm1.6$ &       $1431.7\pm0.8$ &        $978.3\pm0.5$ &        $208.2\pm0.4$ &         $55.1\pm0.4$ &              $<71.5$ & n\\
6 &        $927.5\pm2.6$ &        $316.2\pm0.9$ &              $<81.2$ &     $25979.4\pm69.9$ &              $<81.6$ &        $544.1\pm1.6$ &        $298.4\pm1.5$ &         $71.0\pm1.1$ &  $\cdot \cdot \cdot$ &  $\cdot \cdot \cdot$ & y\\
7 &        $739.1\pm3.1$ &        $251.6\pm1.0$ &        $771.4\pm3.4$ &      $2537.7\pm10.1$ &      $2294.9\pm10.0$ &       $1086.8\pm4.5$ &       $1016.8\pm3.8$ &        $145.2\pm1.0$ &  $\cdot \cdot \cdot$ &  $\cdot \cdot \cdot$ & n\\
10 &         $29.9\pm7.0$ &              $<39.4$ &        $82.4\pm13.7$ &       $229.5\pm34.2$ &       $244.9\pm40.2$ &       $101.4\pm16.9$ &        $77.9\pm12.0$ &         $35.2\pm5.4$ &  $\cdot \cdot \cdot$ &              $<49.2$ & y\\
13 &        $834.3\pm0.2$ &        $295.4\pm0.1$ &       $2263.0\pm0.4$ &       $7168.3\pm1.3$ &       $6870.7\pm1.2$ &       $1395.2\pm0.2$ &       $2096.4\pm0.4$ &        $447.2\pm0.1$ &  $\cdot \cdot \cdot$ &  $\cdot \cdot \cdot$ & n\\
14 &       $212.7\pm11.2$ &         $72.2\pm3.7$ &       $406.2\pm17.5$ &      $1339.3\pm67.6$ &      $1206.3\pm51.5$ &       $278.6\pm14.8$ &       $393.4\pm17.5$ &              $<49.0$ &  $\cdot \cdot \cdot$ &  $\cdot \cdot \cdot$ & y\\
\enddata
\tablenotetext{a}{Swift-BAT 70-month hard X-ray survey ID (\url{http://swift.gsfc.nasa.gov/results/bs70mon/}).}
\tablenotetext{b}{Emission line flux in units of $10^{-17}~{\rm erg}~{\rm cm}^{-2}~{\rm s}^{-1}$.}
\tablenotetext{c}{Flag indicating the use of broad Balmer components (${\rm H\delta, H\gamma, H\beta,~and~H\alpha}$) in spectral line fitting.}
\tablenotetext{d}{Symbols `$\cdot \cdot \cdot$' and `$<$' indicate a lack of spectral coverage and the $3\sigma$ upper limit estimation, respectively.}
\tablecomments{(This table is available in its entirety in a machine-readable form in the online journal. A portion is shown here for guidance regarding its form and content.)}
\label{tab:Hacomplex}
\end{deluxetable*}
\end{rotatetable*}

\begin{rotatetable*}
\begin{deluxetable*}{lcc ccccc cccc}
\tabletypesize{\scriptsize}
\tablecaption{Emission-line Measurements $\textendash$ from [SXII]~$\lambda$7611 to Pa13~$\lambda$8665} 
\tablecolumns{12}
\tablehead{
\colhead{ID\tablenotemark{a}} & 
\colhead{[SXII]\tablenotemark{b}} & 
\colhead{[ArIII]\tablenotemark{b}}  &
\colhead{HeI\tablenotemark{b}} &
\colhead{ArI\tablenotemark{b}} &
\colhead{[FeXI]\tablenotemark{b}} &
\colhead{HeII\tablenotemark{b}} &
\colhead{OI\tablenotemark{b}} &
\colhead{Pa16\tablenotemark{b}} &
\colhead{Pa15\tablenotemark{b}} &
\colhead{Pa14\tablenotemark{b}} &
\colhead{Pa13\tablenotemark{b}} \\
\colhead{} &
\colhead{$\lambda$7611} &
\colhead{$\lambda$7751} &
\colhead{$\lambda$7816} &
\colhead{$\lambda$7868} &
\colhead{$\lambda$7891} &
\colhead{$\lambda$8236} &
\colhead{$\lambda$8446} &
\colhead{$\lambda$8502} &
\colhead{$\lambda$8545} &
\colhead{$\lambda$8598} &
\colhead{$\lambda$8665} 
}
\startdata
     1 &  $\cdot \cdot \cdot$\tablenotemark{c}  &  $\cdot \cdot \cdot$  &  $\cdot \cdot \cdot$  &  $\cdot \cdot \cdot$  &  $\cdot \cdot \cdot$  &  $\cdot \cdot \cdot$  &  $\cdot \cdot \cdot$  &         $24.0\pm0.4$  &  $\cdot \cdot \cdot$  &  $\cdot \cdot \cdot$  &  $\cdot \cdot \cdot$ \\
 2 &  $\cdot \cdot \cdot$  &  $\cdot \cdot \cdot$  &  $\cdot \cdot \cdot$  &  $\cdot \cdot \cdot$  &  $\cdot \cdot \cdot$  &  $\cdot \cdot \cdot$  &  $\cdot \cdot \cdot$  &  $\cdot \cdot \cdot$  &  $\cdot \cdot \cdot$  &  $\cdot \cdot \cdot$  &  $\cdot \cdot \cdot$ \\
 3 &  $\cdot \cdot \cdot$  &  $\cdot \cdot \cdot$  &  $\cdot \cdot \cdot$  &  $\cdot \cdot \cdot$  &  $\cdot \cdot \cdot$  &  $\cdot \cdot \cdot$  &  $\cdot \cdot \cdot$  &        $256.7\pm8.9$  &  $\cdot \cdot \cdot$  &  $\cdot \cdot \cdot$  &  $\cdot \cdot \cdot$ \\
 4 &  $\cdot \cdot \cdot$  &  $\cdot \cdot \cdot$  &  $\cdot \cdot \cdot$  &  $\cdot \cdot \cdot$  &  $\cdot \cdot \cdot$  &  $\cdot \cdot \cdot$  &  $\cdot \cdot \cdot$  &         $62.1\pm0.4$  &  $\cdot \cdot \cdot$  &  $\cdot \cdot \cdot$  &  $\cdot \cdot \cdot$ \\
 5 &        $100.1\pm0.5$  &             $<222.4$  &             $<195.3$  &             $<196.6$  &             $<197.2$  &             $<205.8$  &  $\cdot \cdot \cdot$  &        $208.2\pm0.4$  &  $\cdot \cdot \cdot$  &  $\cdot \cdot \cdot$  &  $\cdot \cdot \cdot$ \\
 6 &  $\cdot \cdot \cdot$  &  $\cdot \cdot \cdot$  &  $\cdot \cdot \cdot$  &  $\cdot \cdot \cdot$  &  $\cdot \cdot \cdot$  &  $\cdot \cdot \cdot$  &  $\cdot \cdot \cdot$  &         $71.0\pm1.1$  &  $\cdot \cdot \cdot$  &  $\cdot \cdot \cdot$  &  $\cdot \cdot \cdot$ \\
 7 &  $\cdot \cdot \cdot$  &  $\cdot \cdot \cdot$  &  $\cdot \cdot \cdot$  &  $\cdot \cdot \cdot$  &  $\cdot \cdot \cdot$  &  $\cdot \cdot \cdot$  &  $\cdot \cdot \cdot$  &        $145.2\pm1.0$  &  $\cdot \cdot \cdot$  &  $\cdot \cdot \cdot$  &  $\cdot \cdot \cdot$ \\
10 &  $\cdot \cdot \cdot$  &              $<47.4$  &  $\cdot \cdot \cdot$  &  $\cdot \cdot \cdot$  &              $<48.2$  &              $<50.3$  &              $<51.6$  &         $35.2\pm5.4$  &              $<52.2$  &              $<52.5$  &              $<62.0$ \\
13 &  $\cdot \cdot \cdot$  &  $\cdot \cdot \cdot$  &  $\cdot \cdot \cdot$  &  $\cdot \cdot \cdot$  &  $\cdot \cdot \cdot$  &  $\cdot \cdot \cdot$  &  $\cdot \cdot \cdot$  &        $447.2\pm0.1$  &  $\cdot \cdot \cdot$  &  $\cdot \cdot \cdot$  &  $\cdot \cdot \cdot$ \\
 14 &  $\cdot \cdot \cdot$  &  $\cdot \cdot \cdot$  &  $\cdot \cdot \cdot$  &  $\cdot \cdot \cdot$  &  $\cdot \cdot \cdot$  &  $\cdot \cdot \cdot$  &  $\cdot \cdot \cdot$  &              $<49.0$  &  $\cdot \cdot \cdot$  &  $\cdot \cdot \cdot$  &  $\cdot \cdot \cdot$ 
 \enddata
\tablenotetext{a}{Swift-BAT 70-month hard X-ray survey ID (\url{http://swift.gsfc.nasa.gov/results/bs70mon/}).}
\tablenotetext{b}{Emission line flux in units of $10^{-17}~{\rm erg}~{\rm cm}^{-2}~{\rm s}^{-1}$.}
\tablenotetext{c}{Symbols `$\cdot \cdot \cdot$' and `$<$' indicate a lack of spectral coverage and the $3\sigma$ upper limit estimation, respectively.}
\tablecomments{(This table is available in its entirety in a machine-readable form in the online journal. A portion is shown here for guidance regarding its form and content.)}
\label{tab:SXIIcomplex}
\end{deluxetable*}
\end{rotatetable*}

\begin{rotatetable*}
\begin{deluxetable*}{lcc ccccc ccccc}
\tabletypesize{\scriptsize}
\tablecaption{Emission-line Measurements $\textendash$ from Pa12~$\lambda$8750 to [SVIII]~$\lambda$9913} 
\tablecolumns{12}
\tablehead{
\colhead{ID\tablenotemark{a}} & 
\colhead{Pa12\tablenotemark{b}} & 
\colhead{[SIII]\tablenotemark{b}}  &
\colhead{Pa11\tablenotemark{b}} &
\colhead{[FeIII]\tablenotemark{b}} &
\colhead{Pa10\tablenotemark{b}} &
\colhead{[SIII]\tablenotemark{b}} &
\colhead{Pa9\tablenotemark{b}} &
\colhead{[SIII]\tablenotemark{b}} &
\colhead{Pa$\epsilon$\tablenotemark{b}} &
\colhead{[CI]\tablenotemark{b}} &
\colhead{[CI]\tablenotemark{b}} &
\colhead{[SVIII]\tablenotemark{b}} \\
\colhead{} &
\colhead{$\lambda$8750} &
\colhead{$\lambda$8829} &
\colhead{$\lambda$8862} &
\colhead{$\lambda$8891} &
\colhead{$\lambda$9014} &
\colhead{$\lambda$9068} &
\colhead{$\lambda$9229} &
\colhead{$\lambda$9531} &
\colhead{$\lambda$9545} &
\colhead{$\lambda$9824} &
\colhead{$\lambda$9850} &
\colhead{$\lambda$9913} 
}
\startdata
   1 &  $\cdot \cdot \cdot$\tablenotemark{c}  &  $\cdot \cdot \cdot$  &  $\cdot \cdot \cdot$  &  $\cdot \cdot \cdot$  &  $\cdot \cdot \cdot$  &  $\cdot \cdot \cdot$  &  $\cdot \cdot \cdot$  &  $\cdot \cdot \cdot$  &  $\cdot \cdot \cdot$  &  $\cdot \cdot \cdot$  &  $\cdot \cdot \cdot$  &  $\cdot \cdot \cdot$ \\
 2 &  $\cdot \cdot \cdot$  &  $\cdot \cdot \cdot$  &  $\cdot \cdot \cdot$  &  $\cdot \cdot \cdot$  &  $\cdot \cdot \cdot$  &  $\cdot \cdot \cdot$  &  $\cdot \cdot \cdot$  &  $\cdot \cdot \cdot$  &  $\cdot \cdot \cdot$  &  $\cdot \cdot \cdot$  &  $\cdot \cdot \cdot$  &  $\cdot \cdot \cdot$ \\
 3 &  $\cdot \cdot \cdot$  &  $\cdot \cdot \cdot$  &  $\cdot \cdot \cdot$  &  $\cdot \cdot \cdot$  &  $\cdot \cdot \cdot$  &  $\cdot \cdot \cdot$  &  $\cdot \cdot \cdot$  &  $\cdot \cdot \cdot$  &  $\cdot \cdot \cdot$  &  $\cdot \cdot \cdot$  &  $\cdot \cdot \cdot$  &  $\cdot \cdot \cdot$ \\
 4 &  $\cdot \cdot \cdot$  &  $\cdot \cdot \cdot$  &  $\cdot \cdot \cdot$  &  $\cdot \cdot \cdot$  &  $\cdot \cdot \cdot$  &  $\cdot \cdot \cdot$  &  $\cdot \cdot \cdot$  &  $\cdot \cdot \cdot$  &  $\cdot \cdot \cdot$  &  $\cdot \cdot \cdot$  &  $\cdot \cdot \cdot$  &  $\cdot \cdot \cdot$ \\
 5 &        $263.5\pm0.9$  &  $\cdot \cdot \cdot$  &  $\cdot \cdot \cdot$  &  $\cdot \cdot \cdot$  &  $\cdot \cdot \cdot$  &        $797.0\pm0.9$  &  $\cdot \cdot \cdot$  &       $1848.5\pm7.9$  &        $162.0\pm7.6$  &  $\cdot \cdot \cdot$  &  $\cdot \cdot \cdot$  &  $\cdot \cdot \cdot$ \\
 6 &  $\cdot \cdot \cdot$  &  $\cdot \cdot \cdot$  &  $\cdot \cdot \cdot$  &  $\cdot \cdot \cdot$  &  $\cdot \cdot \cdot$  &  $\cdot \cdot \cdot$  &  $\cdot \cdot \cdot$  &  $\cdot \cdot \cdot$  &  $\cdot \cdot \cdot$  &  $\cdot \cdot \cdot$  &  $\cdot \cdot \cdot$  &  $\cdot \cdot \cdot$ \\
 7 &  $\cdot \cdot \cdot$  &  $\cdot \cdot \cdot$  &  $\cdot \cdot \cdot$  &  $\cdot \cdot \cdot$  &  $\cdot \cdot \cdot$  &  $\cdot \cdot \cdot$  &  $\cdot \cdot \cdot$  &  $\cdot \cdot \cdot$  &  $\cdot \cdot \cdot$  &  $\cdot \cdot \cdot$  &  $\cdot \cdot \cdot$  &  $\cdot \cdot \cdot$ \\
10 &              $<62.6$  &              $<63.1$  &              $<63.4$  &              $<63.6$  &              $<64.5$  &       $118.3\pm17.1$  &              $<61.1$  &        $61.5\pm11.9$  &  $\cdot \cdot \cdot$  &  $\cdot \cdot \cdot$  &  $\cdot \cdot \cdot$  &  $\cdot \cdot \cdot$ \\
13 &  $\cdot \cdot \cdot$  &  $\cdot \cdot \cdot$  &  $\cdot \cdot \cdot$  &  $\cdot \cdot \cdot$  &  $\cdot \cdot \cdot$  &  $\cdot \cdot \cdot$  &  $\cdot \cdot \cdot$  &  $\cdot \cdot \cdot$  &  $\cdot \cdot \cdot$  &  $\cdot \cdot \cdot$  &  $\cdot \cdot \cdot$  &  $\cdot \cdot \cdot$ \\
 14 &  $\cdot \cdot \cdot$  &  $\cdot \cdot \cdot$  &  $\cdot \cdot \cdot$  &  $\cdot \cdot \cdot$  &  $\cdot \cdot \cdot$  &  $\cdot \cdot \cdot$  &  $\cdot \cdot \cdot$  &  $\cdot \cdot \cdot$  &  $\cdot \cdot \cdot$  &  $\cdot \cdot \cdot$  &  $\cdot \cdot \cdot$  &  $\cdot \cdot \cdot$ 
 \enddata
\tablenotetext{a}{Swift-BAT 70-month hard X-ray survey ID (\url{http://swift.gsfc.nasa.gov/results/bs70mon/}).}
\tablenotetext{b}{Emission line flux in units of $10^{-17}~{\rm erg}~{\rm cm}^{-2}~{\rm s}^{-1}$.}
\tablenotetext{c}{Symbols `$\cdot \cdot \cdot$' and `$<$' indicate a lack of spectral coverage and the $3\sigma$ upper limit estimation, respectively.}
\tablecomments{(This table is available in its entirety in a machine-readable form in the online journal. A portion is shown here for guidance regarding its form and content.)}
\label{tab:SIIIcomplex}
\end{deluxetable*}
\end{rotatetable*}

%

\vspace{5mm}
\facilities{Du Pont (Boller \& Chivens spectrograph), Hale (Doublespec), Keck:I (LRIS), SOAR (Goodman), Swift(BAT), VLT:Kueyen (X-Shooter)}


\software{gandalf \citep{Sarzi06}, ESO Reflex \citep{Freudling13}}



\appendix

\section{Spectral fits}
We provide optical spectral fits of entire BAT AGNs used in this study in the BASS Website (\url{http://www.bass-survey.com}) as shown in Figure~\ref{fig:fits}. Figure~\ref{fig:fits} illustrates achieved spectral fits with an optical image from the Digitized Sky Survey and basic informations.  Spectral fits are displayed with black (observed), red (the best fit), blue (Gaussian narrow components), and green (Gaussian broad components). 

\renewcommand{\thefigure}{A\arabic{figure}}
\setcounter{figure}{0}

\begin{sidewaysfigure}
\centering
\hspace{-46mm}
\includegraphics[width=0.62\textwidth, angle=270]{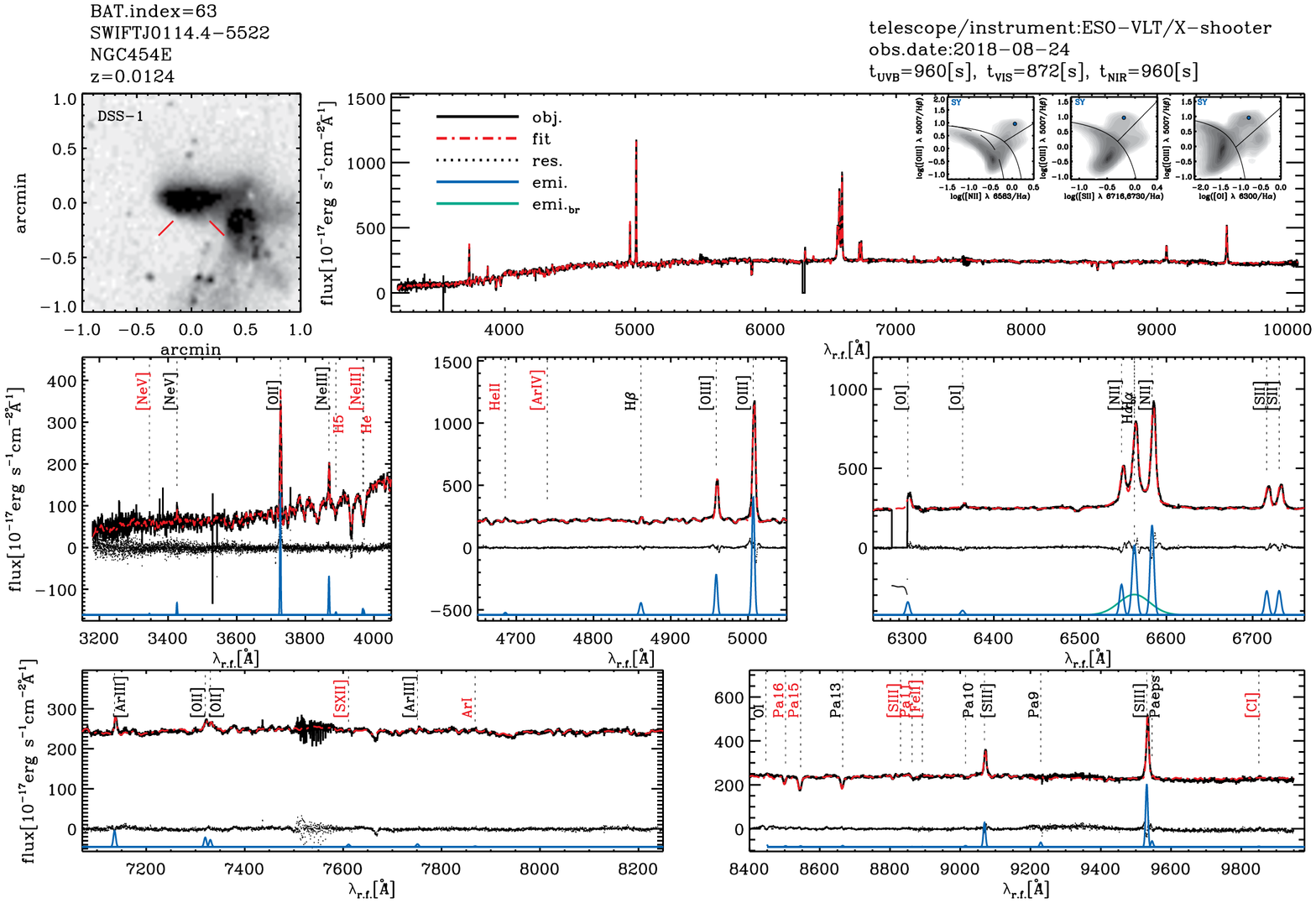}
\caption{Example of spectral line fitting. Counterpart  name, BAT name, redshift, instrument, observed date, and exposure time are shown at the top. Top left: Digitized Sky Survey image in the arcmin scale. Top right: The black line represents the observed spectrum in the rest frame. The red dashed-dotted line is the best fit. BPT diagnostics diagrams are shown in the insets. Middle and bottom rows show the spectral fitting result in detail, and they include the labels of the detected emission-lines. In the case of low A/N, smaller than 3, red labels are used. The blue and green Gaussians are narrow and broad emission-line components, respectively. Residuals are shown in black dots.}
    \label{fig:fits}
\end{sidewaysfigure}




\bibliography{references}{}
\bibliographystyle{aasjournal}



\end{document}